\documentclass{elsart}
\usepackage{graphicx,amssymb,times}
\def\la{\ \raise.3ex\hbox{$<$\kern-.75em\lower1ex\hbox{$\sim$}}\ }
\def\ga{\ \raise.3ex\hbox{$>$\kern-.75em\lower1ex\hbox{$\sim$}}\ }

\begin{document}

\begin{frontmatter}

\title{On the transition from Galactic to extragalactic cosmic-rays: spectral and composition features from two opposite scenarios}

\author[Chicago1,Chicago2]{D. Allard}
\author[IPN]{E. Parizot}
\author[Chicago1,Chicago2]{A. V. Olinto}

\address[Chicago1]{Kavli Institute of Cosmological Physics, University of Chicago, 5640 S. Ellis, Chicago, IL60637, USA.}
\address[Chicago2]{Department of Astronomy and Astrophysics, University of Chicago, 5640 S. Ellis, Chicago, IL60637, USA.}
\address[IPN]{Institut de Physique Nucl\'eaire d'Orsay, IN2P3-CNRS/Universit\'e Paris-Sud, 91406 Orsay Cedex, France.}

\begin{abstract}

We study the phenomenology of cosmic-rays (CRs) at the galactic/extragalactic transition, focusing on two opposite models for the composition of the extragalactic (EG) component. Model A assumes a mixed source composition, with nuclear abundances similar to that of the low-energy CRs, while model B assumes that EG sources accelerate only protons. We study the limits within which both scenarios can reproduce the observed high-energy CR spectrum and composition. The ankle in model A is interpreted as the GCR/EGCR transition, while in model B it is the pair-production dip. Model A has a source spectrum $\propto E^{-x}$ with $x \sim 2.2 - 2.3$, while model B requires $x \sim 2.6 - 2.7$. We compare the predictions of both models with the available data on the energy evolution of the high-energy CR composition using the two main composition-related observables: $X_{\max}$ and $\langle\ln A\rangle$. We conclude that model~A is currently favoured. Uncertainties are discussed and distinctive features of the two models are identified, which should allow one to distinguish between the models in the near future when more precise measurements are available with higher-statistics experiments.

\end{abstract}


\end{frontmatter}

\section{Introduction}
\label{sec:introduction}

Despite its remarkable regularity over more than 12 orders of magnitude in energy, the cosmic-ray (CR) spectrum exhibits a few features that are subject to intense observational and theoretical studies, as they may provide valuable information to eventually understand the origin of both galactic (GCRs) and extragalactic (EGCRs) cosmic-rays. Along with the so-called knee around $5\,10^{15}$~eV, whose precise energy and composition structure is not firmly established yet \cite{Kascade}, and the expected, but observationally uncertain, suppression at high energies due to CR interactions with the cosmological microwave background (CMB), the energy range between $10^{18}$ and $10^{19}$~eV is of particular interest since it is believed to host the transition from GCRs to EGCRs. Such a transition must indeed occur as it is established (notably from observations of Galactic gamma-rays due to $\pi^{0}$ CR interactions) that the low-energy CRs have a Galactic origin, while simple considerations about the confinement of particles in the Galaxy and the Galactic halo strongly suggest that most of the highest-energy CRs must have an extragalactic origin (unless their charge is unexpectedly large, which is not favoured by the observations).

The most natural shape that a transition between two steady (featureless) components may take \emph{a priori} is that of an ``ankle',' i.e., a hardening of the spectrum, where the harder, initially subdominant component at low energy simply takes over from the softer component at \emph{some} transition energy, $E_{\mathrm{t}}$. Conversely, a ``knee-like'' transition corresponding to a steepening of the spectrum with no associated discontinuity is \emph{a priori} unlikely, since it can be smooth only if the second component starts (roughly) at the energy where the first one stops \emph{and} if at that energy they have (roughly) the same flux. Interestingly, an ankle is indeed observed in the CR spectrum, around $3\,10^{18}$~eV, precisely in the energy range where the confining effect of the Galactic magnetic fields is expected to lose its efficiency (because the gyroradius of the particles of charge $Z$ in a microGauss field, $r_{\mathrm{g}} \simeq 1\,\mathrm{kpc}\ Z^{-1}B_{\mu\mathrm{G}}^{-1}$, becomes comparable to the thickness of the Galaxy) and the GCR component is thus expected to die out. It is thus tempting to interpret the observed ankle as a natural signature of the GCR/EGCR transition (e.g., \cite{Nagano92}).

In the $10^{18}$ to $10^{19}$~eV energy range, another feature in the EGCR spectrum is expected, if one assumes that the EGCR sources only accelerate protons. In this case, the energy losses associated with the production of e$^{+}$e$^{-}$ pairs during the proton transport through the CMB lead to a so-called ``pair production dip'' in the propagated spectrum \cite{Berezinsky+88}, the shape of which is reminiscent of the CR ankle. From this point of view, the ankle can be interpreted as a signature of a pure proton EGCR component  \cite{Berezinsky+02,Berezinsky+04}, in which case the GCR/EGCR transition must occur at a lower energy, possibly around the so-called ``second knee'' that may be present in the data but with an uncertain energy scale. Some support for such an interpretation can be found in the recent results of the HiRes experiment \cite{HiRescomp} reporting a rapid variation of the depth of CR-induced extensive air showers (EAS), interpreted as a change of composition from heavy (typically iron) to light nuclei (protons), between $10^{17}$ and $10^{18}$ eV, i.e., below the ankle.
A key assumption in the interpretation of the ankle as a pair production dip is the effective  absence of nuclei heavier than hydrogen among EGCRs, not only by the time they reach the Earth but also at their source \cite{Berezinsky+04,Allard2005}. 

Another important parameter of the EGCR source scenario is the logarithmic slope of the assumed power-law injection spectrum, $Q(E) \propto E^{-x}$. In a recent study \cite{Allard2005}, we have drawn attention to the influence of the EGCR source composition on both the interpretation of the ankle and the acceleration scenario. Our results can be summarized as follows: i) if the sources accelerate only protons, then the observed high-energy CR spectrum can be well reproduced with a steep injection spectrum, $x\simeq 2.6$--2.7 \cite{Berezinsky+02,De marco+03}, and the ankle can be interpreted as a pair-production dip, with a GCR/EGCR transition  $\la 10^{18}$~eV; ii) if the EGCR source composition is roughly similar to that of GCRs, i.e., a substantial faction of heavier nuclei from the ambient medium around the source are accelerated together with protons, then the observed CR spectrum can be reproduced equally well with a less steep injection spectrum, $x\simeq 2.2$--2.3, and the ankle can be interpreted as the GCR/EGCR transition (around $3\,10^{18}$~eV).

While an ankle-like transition is \emph{a priori} more natural than a knee-like transition, the rough coincidence between the end of the GCR component and the beginning of the EGCR component may arise naturally around $10^{18}$~eV, as a result of the shortage of high-energy GCRs (due to either acceleration or propagation effects) and the suppression of low-energy EGCRs because of expansion losses and/or magnetic field filtering \cite{Aloisio2005,Lemoine+05}. Also, while a very steep spectrum and a discrimination of heavy nuclei by the acceleration process is \emph{a priori} less natural, it has been claimed that strong extended magnetic fields around EGCR sources might substantially increase the path of highly charged nuclei compared to protons and thus tend to suppress heavy nuclei, turning an originally mixed composition into an effective pure proton source, with a subsequent phenomenology similar to that of the genuinely pure-proton scenario \cite{Gunter}.

However that may be, we do not focus here on the theoretical plausibility of the different models, which remains somewhat subjective at this stage, but rather investigate the two main scenarios  from the point of view of composition. In particular, we make  definite predictions about observable features in the GCR/EGCR transition region for both models and compare our results with the available data.

In Sect.~2, we summarize the different composition hypotheses and briefly recall our standard scheme for the photo-disintegration of ultra-high-energy nuclei. In Sect.~3, we compare the propagated spectra obtained within the two main scenarios investigated, their implications for the remaining galactic component and the phenomenological limits of both models. We characterize the transition in terms of composition observables in Sect.~4 and compare our results with the available data. Finally, we discuss our results and conclude in Sect.~5.

\section{Physical and astrophysical inputs}
\label{sec:PhysAstrophys}

The effect of the CMB in the propagation of ultra-high-energy (UHE) nuclei in the intergalactic space was first  discussed in the seminal papers by Greisen \cite{G66} and Zatsepin and Kuzmin \cite{ZK66} where the well-known GZK cutoff was predicted. The first detailed study of the propagation of UHE nuclei heavier than protons can be found in \cite{PSB} (see \cite{Stecker99} for an update), who showed that the main limiting process was the interaction of UHE nuclei with the CMB photons through the giant dipolar resonance (GDR). Since the energy threshold for this process roughly scales with the atomic number, $A$, it was realised that only iron (or heavier) nuclei could contribute significantly to the UHECR spectrum above $5\,10^{19}$~eV. Most of the following studies of UHE nuclei propagation then considered only iron \cite{Anchordoqui1998,Epele+98,Bertone+02} or super-heavy nuclei \cite{Anchordoqui1999}, while only a few considered also low and intermediate mass nuclei \cite{Anchordoqui2001}. It should be stressed that although the intermediate mass nuclei do not contribute substantially to the spectrum at the highest energies (at least under the most standard assumptions), their influence can be noticeable in the ankle energy range, which is of particular interest to us here. As recalled above, we have shown recently that including these nuclei in the source composition radically changes  the interpretation of EGCR phenomenology by allowing the CR injection spectrum to be similar to that generally expected from relativistic shock acceleration \cite{Allard2005}.

Since the EGCR composition is essentially unknown, we investigate two typical cases following the above-mentioned plausible scenarios for the GCR/EGCR transition and the interpretation of the ankle. The first scenario is characterized by a mixed composition, with abundance ratios \emph{at the source} similar to those of the much better known low energy CRs (model~A), and the second by a pure proton source (model~B). For model~A we use the values given \cite{Duvernois96}, simply transforming the relative abundances, $\alpha_{i}$, at a given energy per nucleus ($E/A$) into relative abundances at a given energy, $\zeta_{i}=\alpha_{i}A_{i}^{x-1}$, where $x$ is the injection spectral index of the source \cite{Allard2005}. The injection spectrum of each nuclear species, $i$, is then given by $N_{i}(E)\propto \alpha_{i}A_{i}^{x-1}E^{-x}$ which we refer to as our ``generic composition'' for the mixed-composition scenario. This scenario has a mild dependence on the EGCR spectral index, but our results are found not to depend significantly on the details of the composition, at least within reasonable limits, as discussed below. Likewise we consider some departure from the pure proton case by adding small fractions of He or Fe nuclei, and study the validity domain of the pair-production dip interpretation of the ankle within this scenario.

Concerning the high-energy end of the injection spectrum, we assume a simple exponential cut-off at some energy $E_{\max}$, and we assume that the acceleration mechanism is governed by rigidity-dependent processes, such that the maximum energy of a nucleus of charge $Z_{i}$ is $Z_{i}$ times higher than that of protons: $E_{\max}(Z_{i})=Z_{i}\ E_{\max}(^{1}H)$. We use the typical value of $E_{\max}(^{1}H)=10^{20.5}$~eV, but study also the influence of this parameter.

The EGCR propagation is calculated by taking into account the main energy loss mechanisms for protons and nuclei 
due to the interaction with the CMB and the infra-red, optical and ultraviolet backgrounds (IR/Opt/UV). For the three latter, we used the latest estimates (for the density and the cosmological evolution) calculated by \cite{MS05}.
In addition to the universal expansion losses and the pair production losses, both treated as continuous processes, we use a stochastic approach to describe the photo-production of pions in the case of protons propagating through the CMB and IR/Opt/UV photon fields, as well as the four main photo-erosion processes for the propagation of UHE nuclei. The giant dipolar resonance (GDR) and the quasi-deuteron processes (QD) were considered in the seminal study of Puget et al. \cite{PSB}. For the GDR (which is the main photo-dissociation process because of its lower energy threshold and larger cross-section), we use the new set of cross-sections recently calculated by \cite{Khan05}, giving a much better description of the experimental data. This allows us to follow the propagation of all the nuclei in the two-dimensional $(N,Z)$ nuclear space. In addition, we consider the processes of baryonic resonances (BR) and photo-fragmentation (PF), which contribute at the highest energies. Although the GDR is the key process for the propagation of UHE nuclei, the BR contribution to the photo-nuclear cross-sections has the important consequence of keeping the nuclei attenuation lengths at their lowest values even at very high energy where the influence of the GDR decreases. This is due to the corresponding large cross-sections and nucleon yields. A description of the relative influence of the four processes on the iron mean free path can be found in \cite{Allard2005,Allard2006}, where we also showed the evolution of the attenuation length as a function of energy. Useful parameterizations of the QD, BR and PF cross-sections can be found in \cite{Rachen}. Further details on our propagation code are given in \cite{Allard2005,Allard2006}.

In the following, we assume that the extragalactic magnetic fields do not strongly influence the propagation of EGCRs, and postpone until Sect.~5  a short discussion about their possible effects. Likewise, to limit the number of free parameters, we make the usual assumption that the EGCR sources are distributed uniformly throughout the universe, with no significant evolution of their properties (spectrum and intensity) over the last few gigayears. Equally good fits leading to a totally similar interpretation of the ankle can however be found with stronger evolution providing a slightly harder injection index (2.1-2.2) to compensate the higher contribution of far sources as in the pure proton case. It is important to note that nuclei heavier than protons are not disfavored by strong source evolution because the redshift evolution of the IR/Opt/UV backgrounds \cite{MS05} is small compared to that of the CMB. The nuclei are thus not affected much by a change of the source evolution.  A more detailed study of the effects induced by a stronger source evolution on the predicted spectrum can be found  in \cite{Allard2006}. We also assume the simplest functional shape for the injection spectrum, namely a power law, $E^{-x}$, where the logarithmic slope, $x$, is a free parameter. It should be noted, however, that in the case of a pure proton source composition, the value of $x$ necessary to fit the data (namely $x = 2.6$--2.7) is so large that one cannot assume such a spectrum to hold down to thermal energies without running into an energy crisis, requiring an unreasonably large source power. For this reason, one must assume that the EGCR injection spectrum somehow gets flatter at low energies, which introduces at least two additional free parameters, namely a break energy, $E_{\mathrm{br}}$, and the slope of the spectrum below that energy. This is further discussed below. Such a complication is not met in the mixed-composition case, since the corresponding injection spectrum is flatter.

\section{Spectral constraints on EGCR models}

In Figs.~\ref{fig:spectraMixedCompo} and~\ref{fig:spectraPureProtons}, we show the propagated spectra multiplied by $E^{3}$ (i.e., $ E^{3} \Phi(E)$) obtained with the two models described above, A and B. The injection spectral index and the source power are adjusted so as to provide a good fit of the high-energy CR data. We use separately the HiRes data \cite{Bergman+05} and the Akeno/AGASA data \cite{Nagano92,AGASA}.  For the latter, we shift the AGASA energy scale down by $10 \%$ to obtain a common normalization with Akeno \cite{NaganoWatson} and to follow the recent re-analysis of AGASA results \cite{newAGASA}. Comparison with the Auger spectrum \cite{Augerspec} gives the same results for the spectral fits and their interpretation, since the Auger spectrum is consistent  with both HiRes and AGASA spectra within statistical and systematic uncertainties. We shall not consider further the Auger data here since no composition results are available yet.

\subsection{Propagated spectra for model A (mixed composition)}

\begin{figure}[t]
\centering
\includegraphics[width=0.48\linewidth]{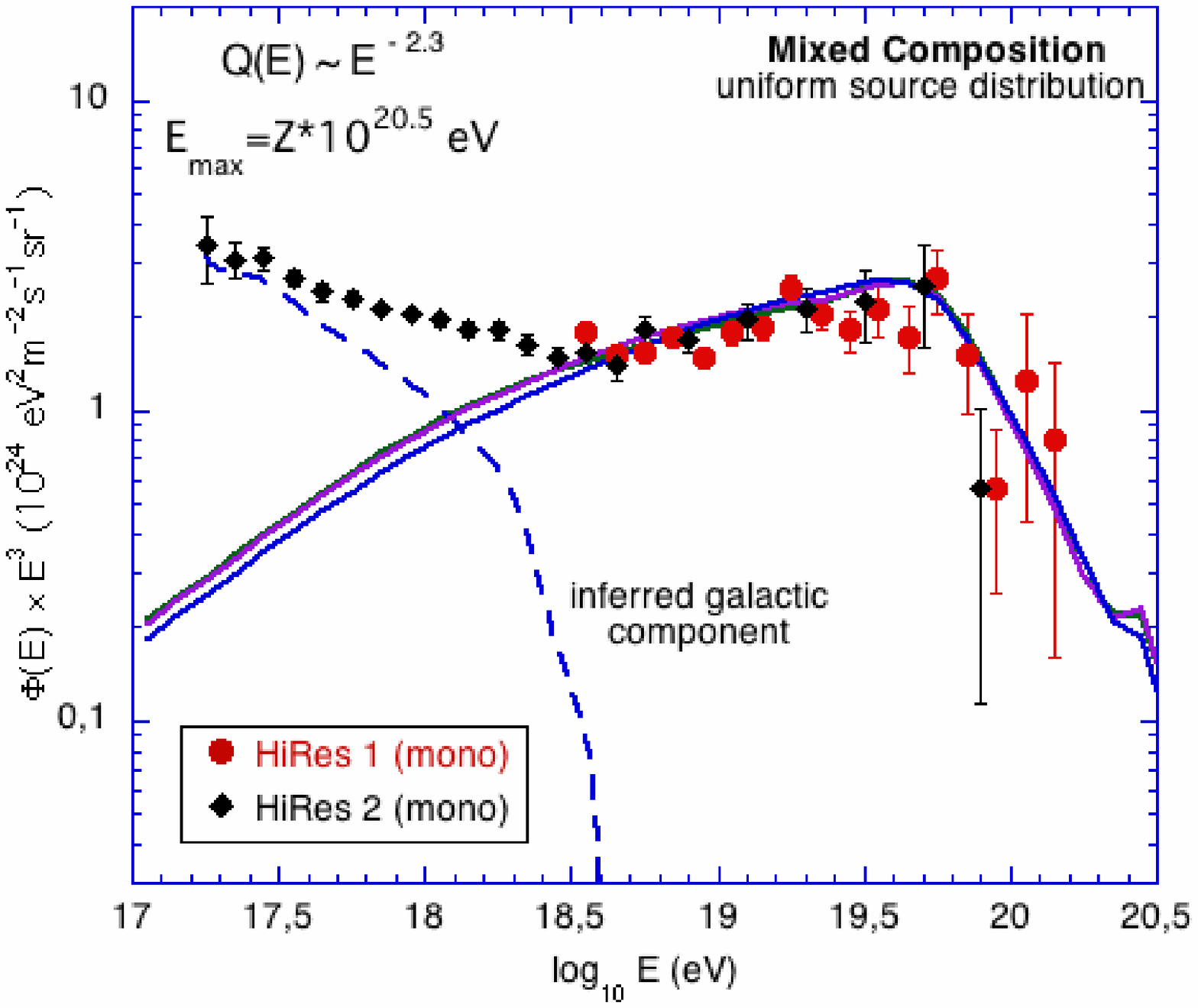} \hfill
\includegraphics[width=0.48\linewidth]{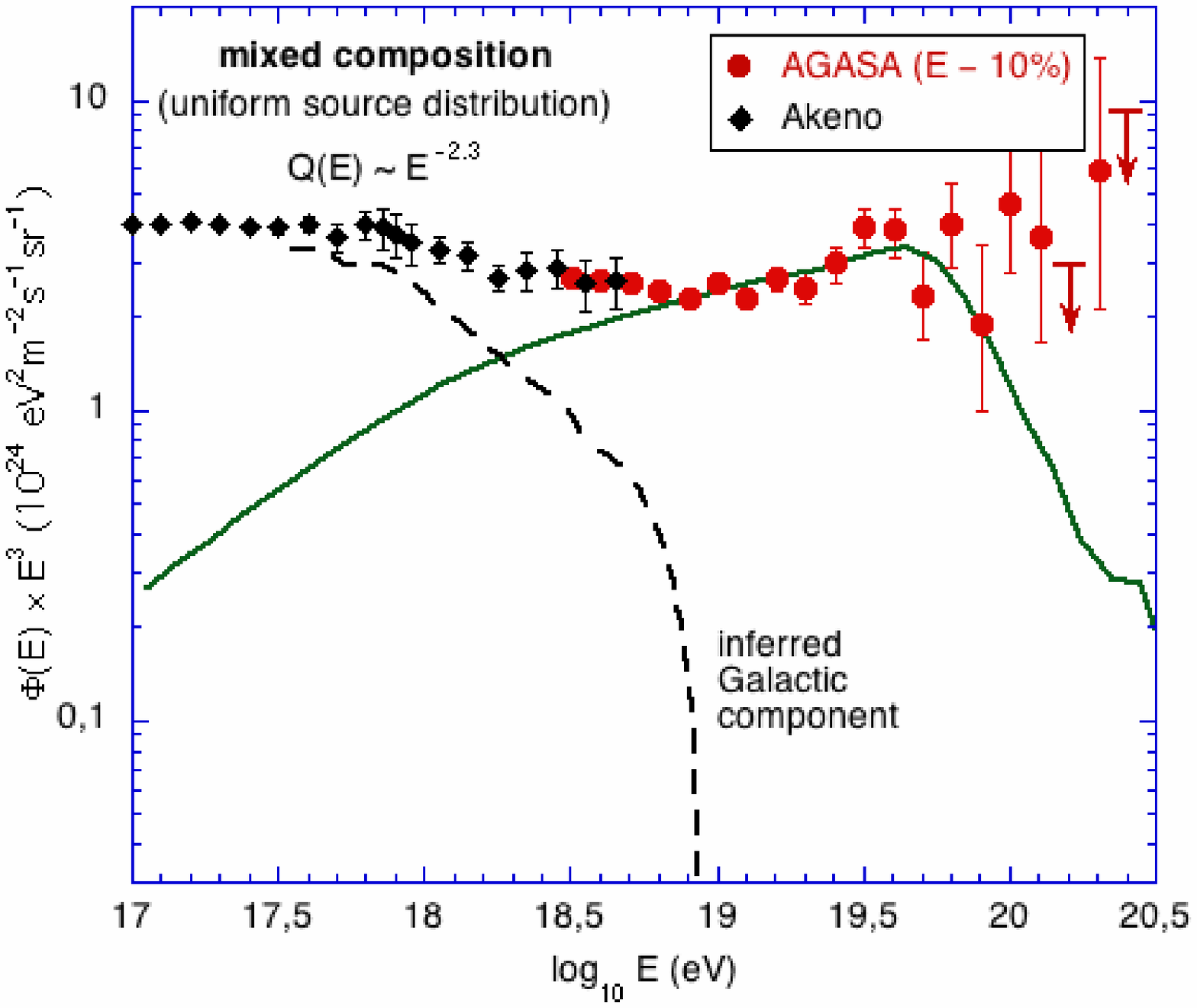}
\caption{Propagated spectra, $\Phi(E) \, E^{3}$, for model A (mixed EGCR source composition), compared with the HiRes data (left) and the Akeno/AGASA data (right). The dashed line corresponds to the inferred GCR component, obtained by subtraction of the EGCR component from the total CR flux (data points). In both cases the source spectrum has a logarithmic index of $x = 2.3$, and the ankle corresponds to the GCR/EGCR transition. On the left, the three propagated spectra correspond to three different compositions (see text), with essentially identical results.}
\label{fig:spectraMixedCompo}
\end{figure}

Figure~\ref{fig:spectraMixedCompo} shows the results of our propagation code for a mixed EGCR source composition (model A). It can be seen in Fig.~\ref{fig:spectraMixedCompo}a that this model provides a good fit of the high-energy HiRes data, with a source spectrum  $Q(E)\propto E^{-2.3}$. Results are shown for three different compositions: i) the standard composition, identical to that of the low-energy cosmic rays (at the source), ii) a modified composition, with two times less He nuclei and two times more CNO nuclei, and iii) a second modified composition, with two times less He nuclei and three times more Fe nuclei (lower curve). As can be seen, the propagated spectra are essentially identical in all three cases, except at low energies where the Fe rich composition gives slightly lower fluxes.

In Fig.~\ref{fig:spectraMixedCompo}b,  the propagated spectrum is compared with the Akeno/AGASA data. As is well know, the highest energy data are only marginally consistent with the predicted \emph{GZK suppression} of the flux, within 2.3 to 2.6 $\sigma$ depending on the energy scale \cite{De marco+03}. If the excess flux at high energies is confirmed by higher statistics observations, it  
 will require an additional component for both model A and model B. Apart from this high-energy tail, the data suggest a source spectrum with a logarithmic slope for model A of $x = 2.3$, as with the HiRes data.

In both cases, we show the inferred Galactic component, obtained by simply subtracting the EGCR component from the observed data. In this scenario, the ankle corresponds to the GCR/EGCR transition \cite{Allard2005}. To be more precise, it marks the end of the transition, where the Galactic flux becomes negligible, which occurs around $3\,10^{18}$~eV in the HiRes data or at $6\,10^{18}$~eV for the Akeno/AGASA data. The GCR flux is totally negligible above $10^{19}$~eV. The transition is seen to take place over about one order of magnitude in energy, with equal fluxes in the GCR and EGCR components around 1.5--2~$10^{18}$~eV.

\subsection{Propagated spectra for model B (protons only)}

\begin{figure}[t]
\centering
\includegraphics[width=0.48\linewidth]{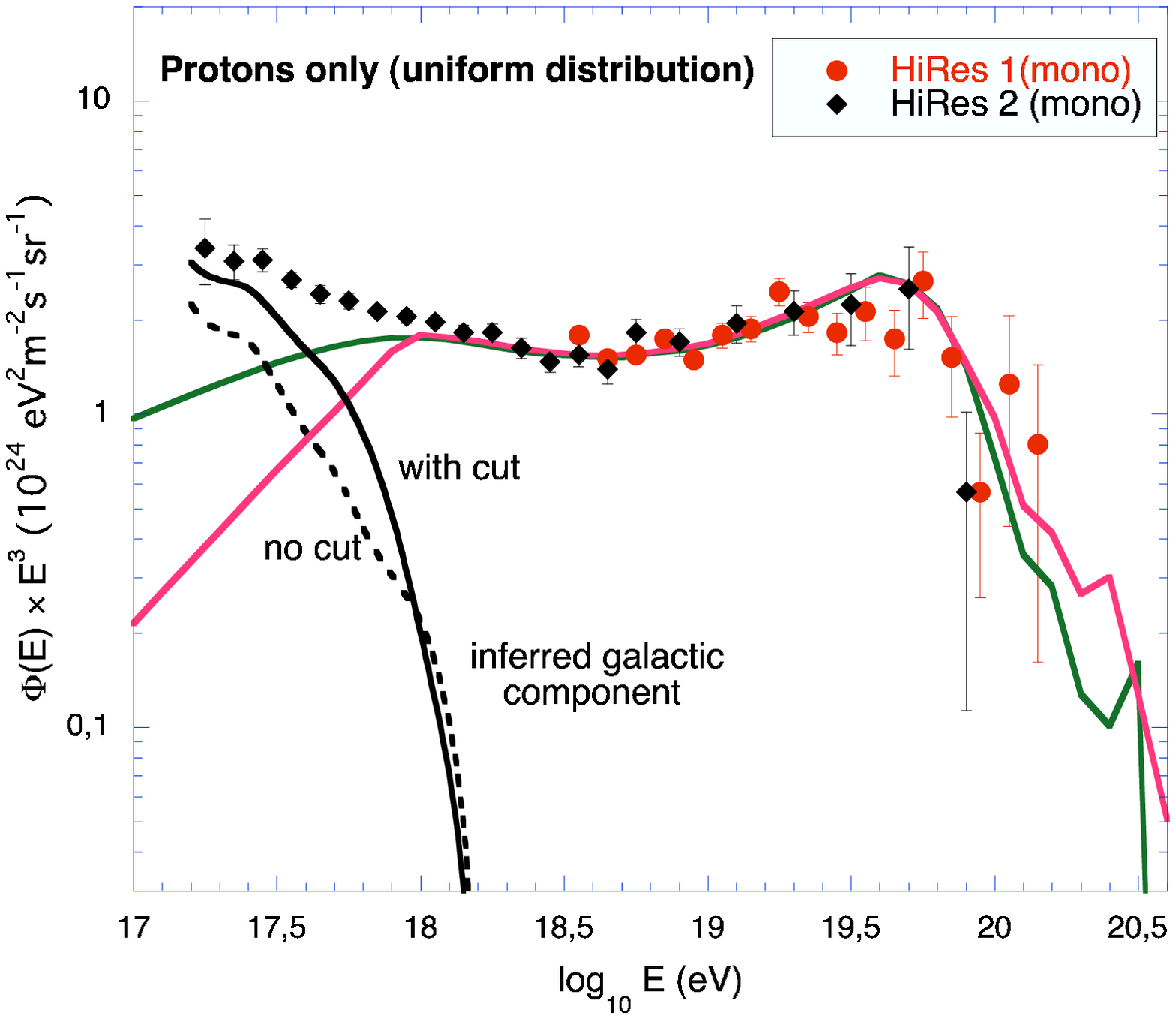} \hfill
\includegraphics[width=0.48\linewidth]{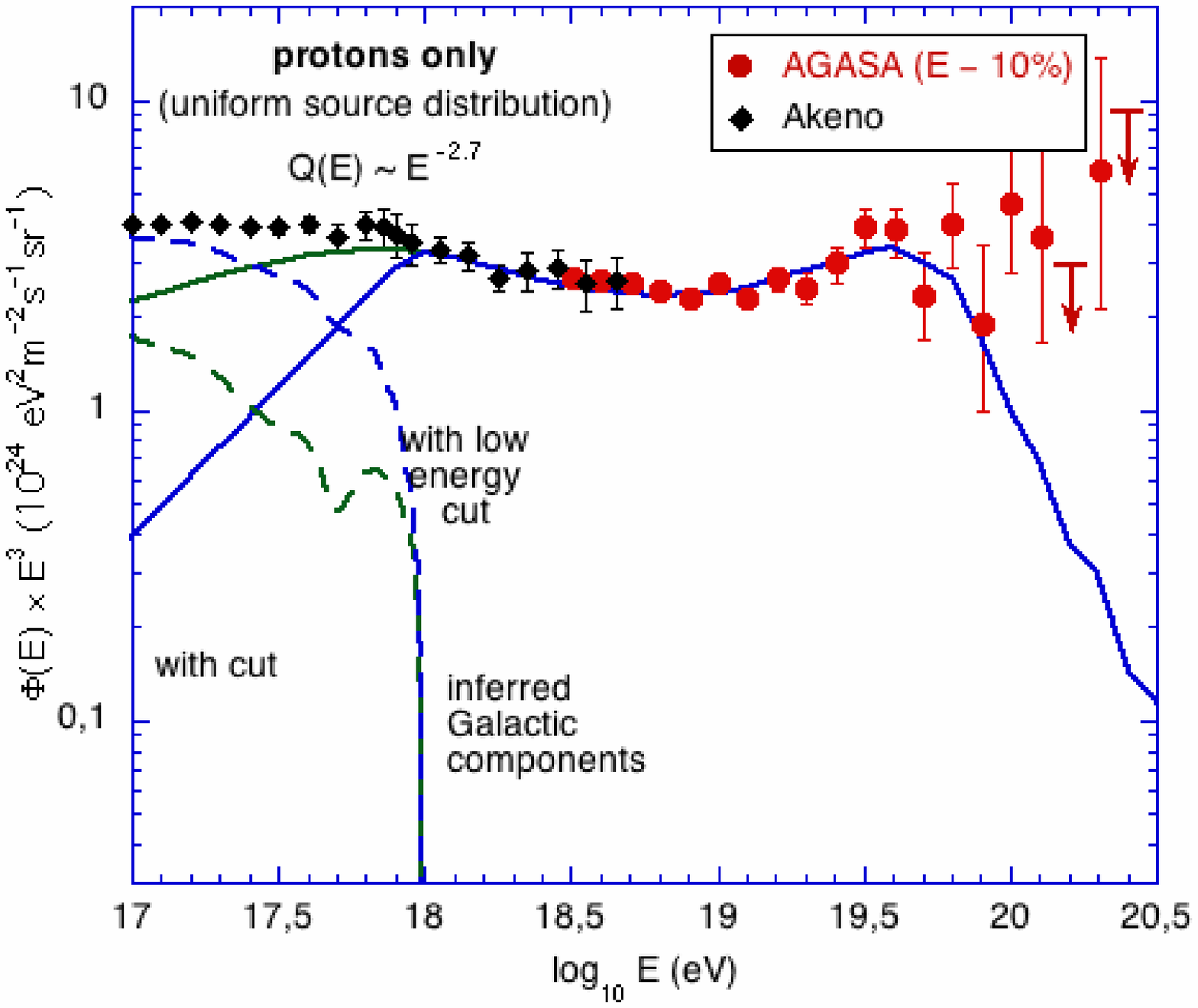}
\caption{Same as Fig.~\ref{fig:spectraMixedCompo}, for model B (pure proton EGCR sources). The injection spectral index is $x = 2.6$ for the HiRes data (left) and $x = 2.7$ for the Akeno/AGASA data (right). Two different propagated spectra and the corresponding inferred GCR component are shown, for an injection spectrum either with or without a low energy cut (see text).}
\label{fig:spectraPureProtons}
\end{figure}

Figure~\ref{fig:spectraPureProtons} shows the results in the case of model B (pure proton EGCR sources). As shown in the figure, a good agreement is obtained with the high-energy data down to $10^{18}$ eV, assuming an injection spectral index of $x = 2.6$ in the case of HiRes (on the left), and $x = 2.7$ for the Akeno/AGASA data (on the right), as found in previous works \cite{Berezinsky+02,De marco+03}. As noted in Sect.~\ref{sec:introduction}, the steep injection spectrum leads to an energy crisis at low energies requiring a change of slope to make the spectrum flatter at low energies \cite{Berezinsky+02}.  

While a break at low-energy is required in the case of model B, the observed spectrum does not allow us to constrain neither the energy, $E_{\mathrm{br}}$, where it occurs nor the slope below $E_{\mathrm{br}}$. In Fig.~\ref{fig:spectraPureProtons}, we show the propagated spectra obtained in two cases: i) without a cut (i.e., implicitly with a cut below $10^{17}$~eV), and ii) with a break in the spectrum at $E_{\mathrm{br}} = 10^{18}$~eV and a spectral index $x = 2.0$ below the $E_{\mathrm{br}}$. In each case, we show the resulting Galactic component necessary to account for the observed fluxes in the transition region. In the first case (no break), the EGCR component appears to dominate down to $3\,10^{17}$~eV if one refers to the HiRes data (Fig.~\ref{fig:spectraPureProtons}a), or even down to below $10^{17}$~eV if one refers to the AGASA data (Fig.~\ref{fig:spectraPureProtons}b). 
Since the EGCR component is made of protons only in this model,  the case of no cut clearly implies a large fraction of protons at $10^{17}$~eV, larger than 20\% for HiRes, or even 50\% for AGASA, which seems to be disfavoured by the recent results of the Kascade experiment \cite{Kascade}. In the second case, when a low-energy cut is imposed on the EGCR injection spectrum, it is clearly possible to reduce the extragalactic contribution to the total CR flux at $10^{17}$~eV. This reduction comes at the expense of a sharper GCR/EGCR transition, as shown in Fig.~\ref{fig:spectraPureProtons}. 

In both cases of model B and for any choice of data sets, the pair-production dip interpretation of the ankle implies that the GCR component has essentially died by $10^{18}$~eV, in contrast with the case of model A. The resulting composition structure at and below the ankle will thus be very different. The aim of this paper is to make definite predictions concerning the composition observables in this region and show how present and future data can be used to discriminate between model A and model B. Note that if a low-energy cut is applied to model B, the transition must occur over an even shorter range of energy, namely less than half an order of magnitude, which implies a very sharp change of the observed CR composition, and thus a very large elongation rate. (The elongation rate, $\mathcal{E}$, is defined as the derivative of $X_{\max}$, the shower maximum, with respect to the logarithm of the energy, $\mathcal{E} = \mathrm{d}X_{\max}/\mathrm{d}\log E_{\mathrm{eV}}$.)

Finally, as mentioned above, sufficiently large extragalactic magnetic fields could lead to an additional suppression of the low-energy EGCRs \cite{Aloisio2005,Lemoine+05,Parizot+05}. However, this suppression would only make the transition sharper and it would mimic a low-energy break at the source even if there is none.

\subsection{Phenomenological limits of model A}

In the absence of a definite source model, the initial composition of the EGCRs cannot be predicted  precisely. As shown above, the data is consistent with a mixed composition similar to that of the better-known low-energy CRs, as would result if the mechanisms by which the particles are injected into the acceleration process are similar in all sources. This behavior is expected for most MHD processes operating after a possibly selective injection correlated with physical parameters such as the volatility or the first ionisation potential of the different elements. Some departure of the EGCR composition from our ``generic composition''  are however expected, but as we showed in Fig.~\ref{fig:spectraMixedCompo}a, our results are quite robust to even significant composition changes. 

\begin{figure}[t]
\centering
\includegraphics[width=0.48\linewidth]{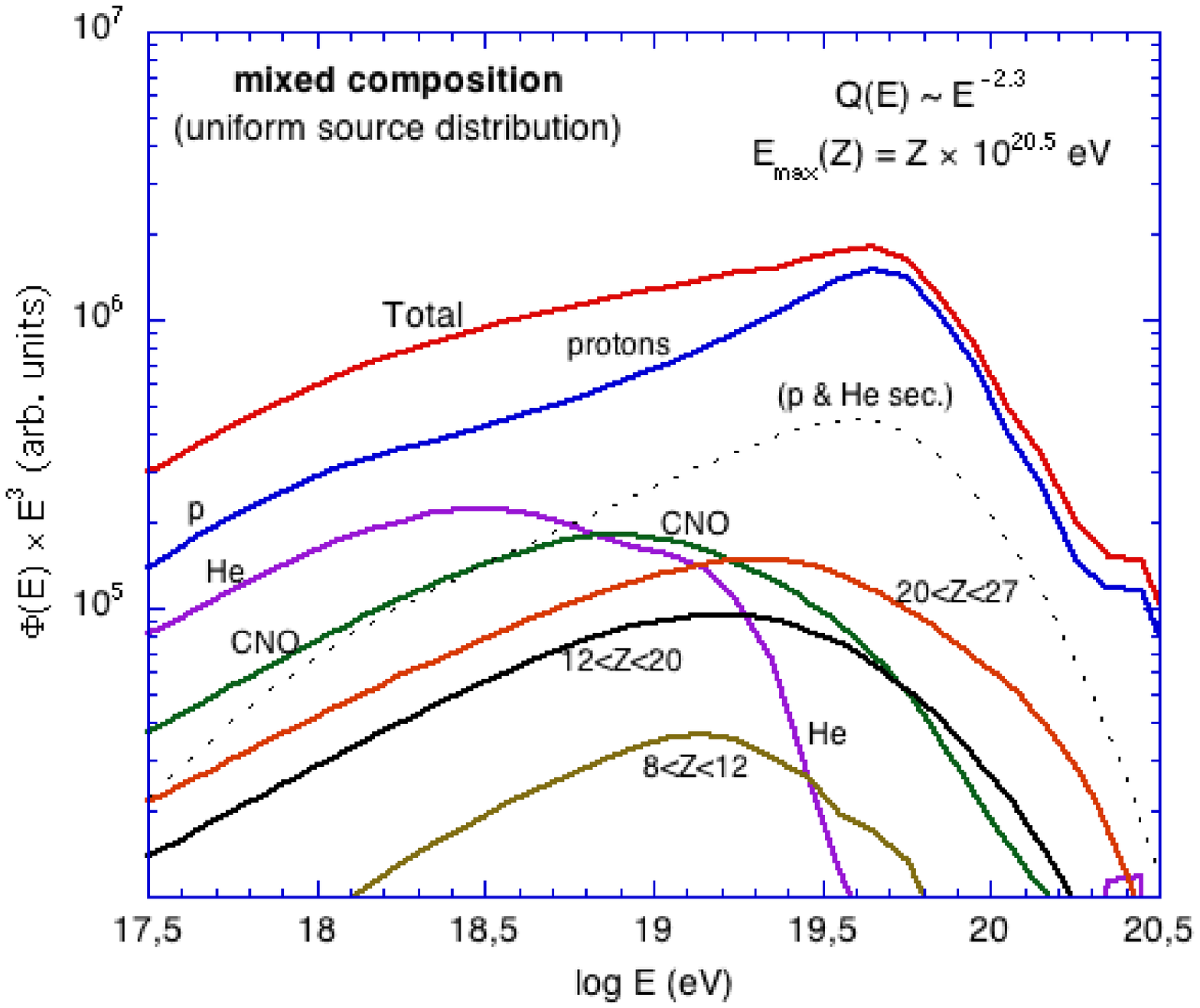} \hfill
\includegraphics[width=0.48\linewidth]{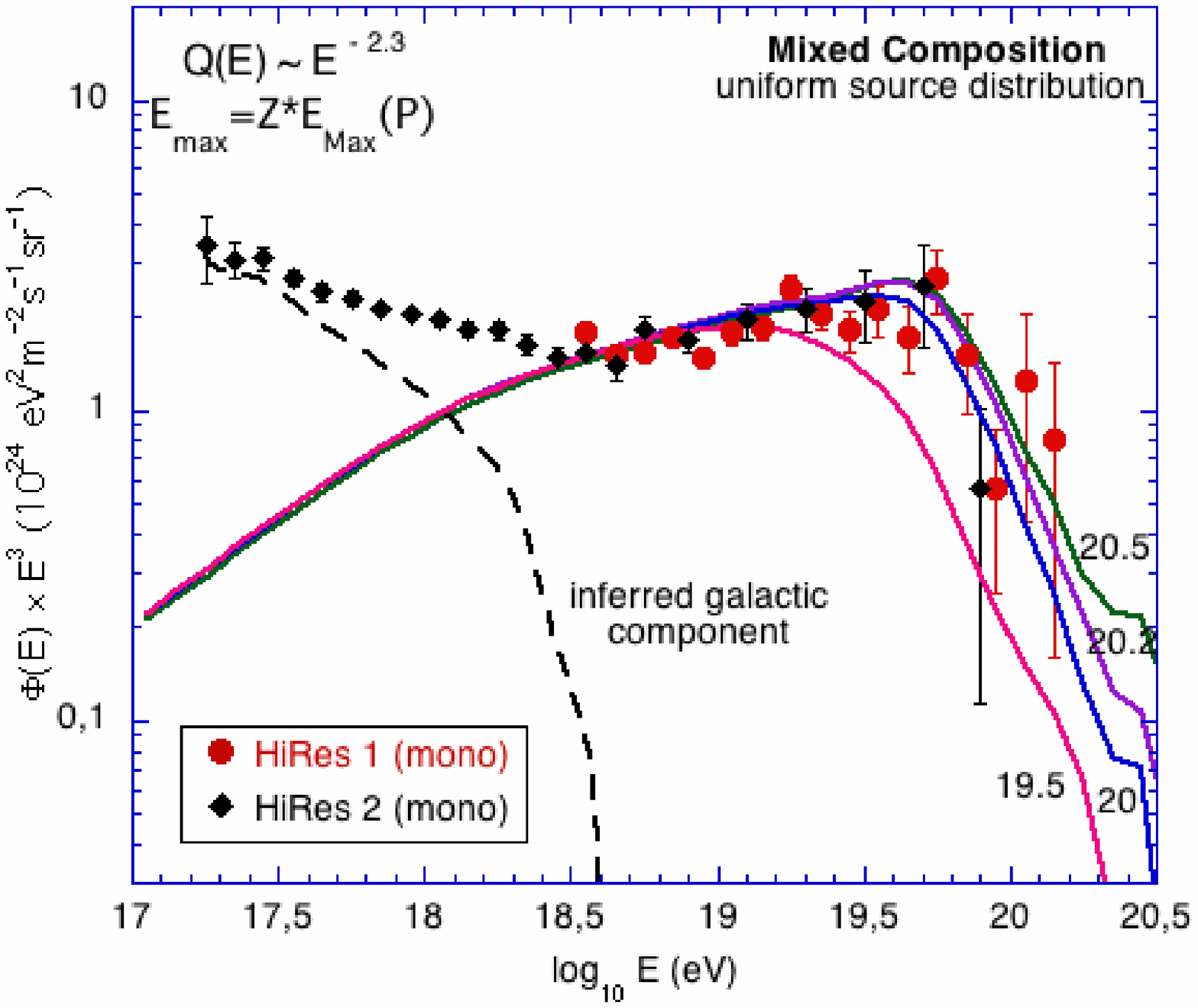}
\caption{On the left  the contribution of different elements or element classes to the propagated spectrum of Fig.~\ref{fig:spectraMixedCompo}b is indicated by the labels. GZK suppressions can be seen around $10^{19}$~eV for He and $2\,10^{19}$~eV for the CNO group. The dotted line shows the contribution of the secondary low mass nuclei (protons and He) resulting from the photo-dissociation of heavier nuclei. Their contribution is responsible for the bump in the spectrum around $5\,10^{19}$~eV. On the right  propagated spectra are shown for different values of the proton maximum energy, $E_{\max}(\mathrm{p}) = 10^{a}$~eV, with $a = 19.5$, 20.0, 20.2 and 20.5, as indicated. A low-energy cutoff of the proton spectrum is inconsistent with the data.}
\label{fig:components}
\end{figure}

To better understand how the propagated spectrum depends on the injection composition, we show the spectrum of each components in Fig.~\ref{fig:components}a. The figure shows how the pair-production dip, which is clear in the pure proton case, is weakened by the contribution of the other elements. This compensation by elements heavier than protons is the reason why a mixed composition leads to a completely different interpretation of the ankle and the GCR/EGCR components.
The figure also shows that He and intermediate mass nuclei (CNO, Mg, Si) have their photo-erosion cutoffs, through interactions with the CMB, between 10 and 40~EeV. As a consequence, if these species were to dominate the EGCR source composition, a significant cutoff in the observed spectrum would result. This essentially sets a lower limit to the fraction of protons at the source.  We shall simply note here that as long as the protons make the dominant contribution to the injection spectrum, the predicted propagated spectra do not significantly depend on the exact details of the remaining composition. Compositions lighter or heavier than the generic one can give an equally accurate fit of the UHECR data.

Apart from lowering the proton abundance at the source, another way to have the He- and intermediate-mass-nuclei cutoffs appear in the propagated spectrum (which would then drop at too low an energy) is to reduce the maximum proton energy, $E_{\max}$(p), at the source. This is shown if Fig.~\ref{fig:components}b, where we plot the propagated spectra obtained in the same conditions as in Fig.~\ref{fig:spectraMixedCompo}b, but with different values of $E_{\max}$(p), namely $10^{a}$~eV, with $a = 19.5$, 20.0, 20.2 and 20.5. As can be seen, a proton maximum energy lower than $10^{20}$~eV would not be consistent with the HiRes data, even though heavier nuclei can still reach higher energies in this case since $E_{\max}(Z) = Z\, E_{\max}(\mathrm{p})$. Clearly, higher statistics are needed at high energies to better constrain $E_{\max}$(p). Finally, although we assumed that the maximum energy of each species is proportional to their charge, it should be noted that once the proton cutoff at the source, $E_{\max}$(p), is fixed, the predicted spectra are rather insensitive to the maximum energy of the other species. 

The rigidity dependent maximum energy is not critical for the viability of the model, but adjustments to the spectral index or proton fraction maybe necessary as one relaxes the rigidity dependent maximum energy assumption. For example,  if the maximum energy reached by nuclei heavier than protons is low (for example, $E_{\max}(Z)  \la Z \ 10^{19}$ eV, for $Z>1$ due to photo-dissociation at the source), no high energy secondaries are produced leading to the flattening of the spectrum at high energies. In this case a slightly higher proton fraction than the 40\% assumed or a harder injection spectrum would be required to fit the observed spectra.

\begin{figure}[t]
\centering
\includegraphics[width=0.48\linewidth]{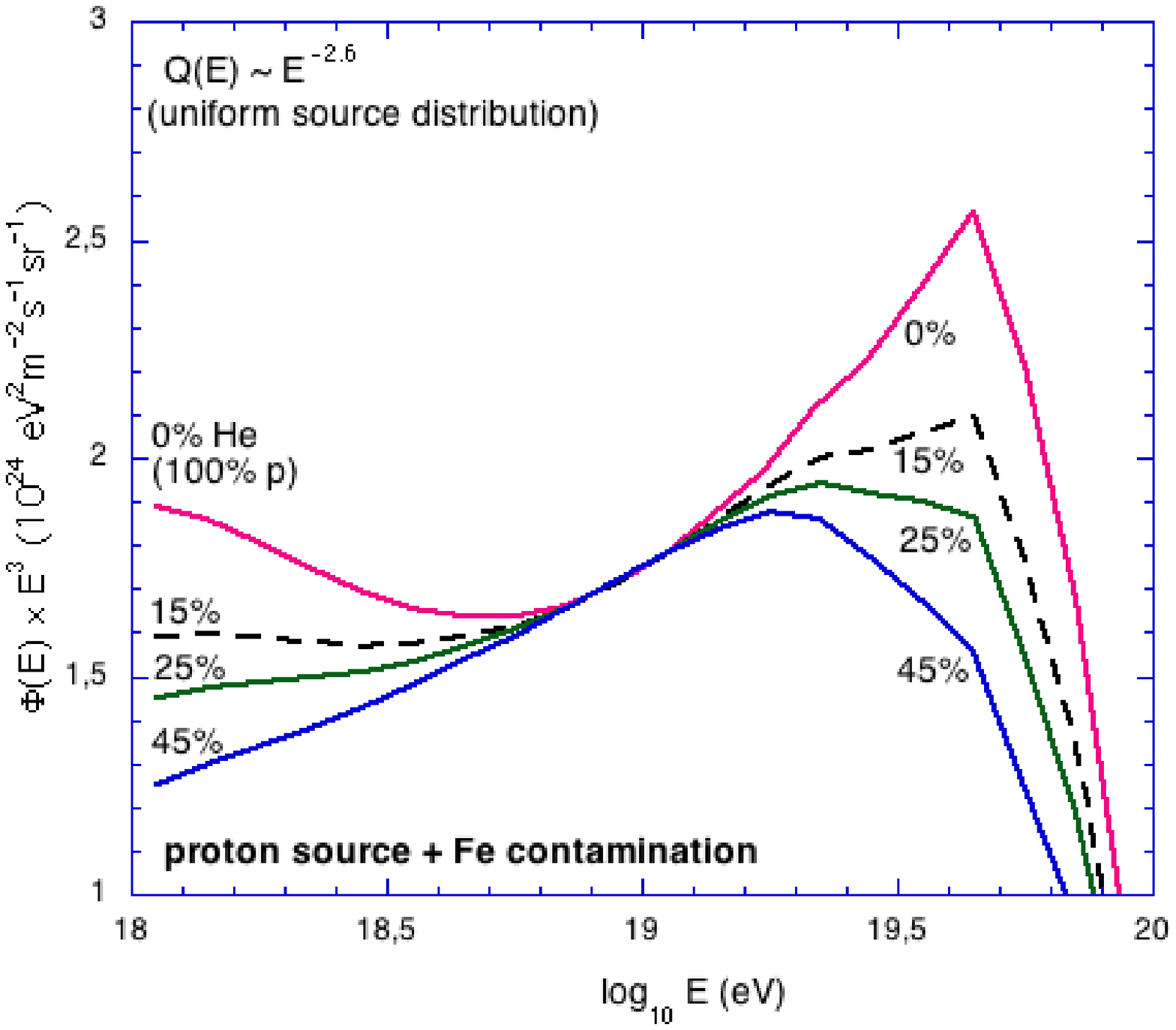} \hfill
\includegraphics[width=0.48\linewidth]{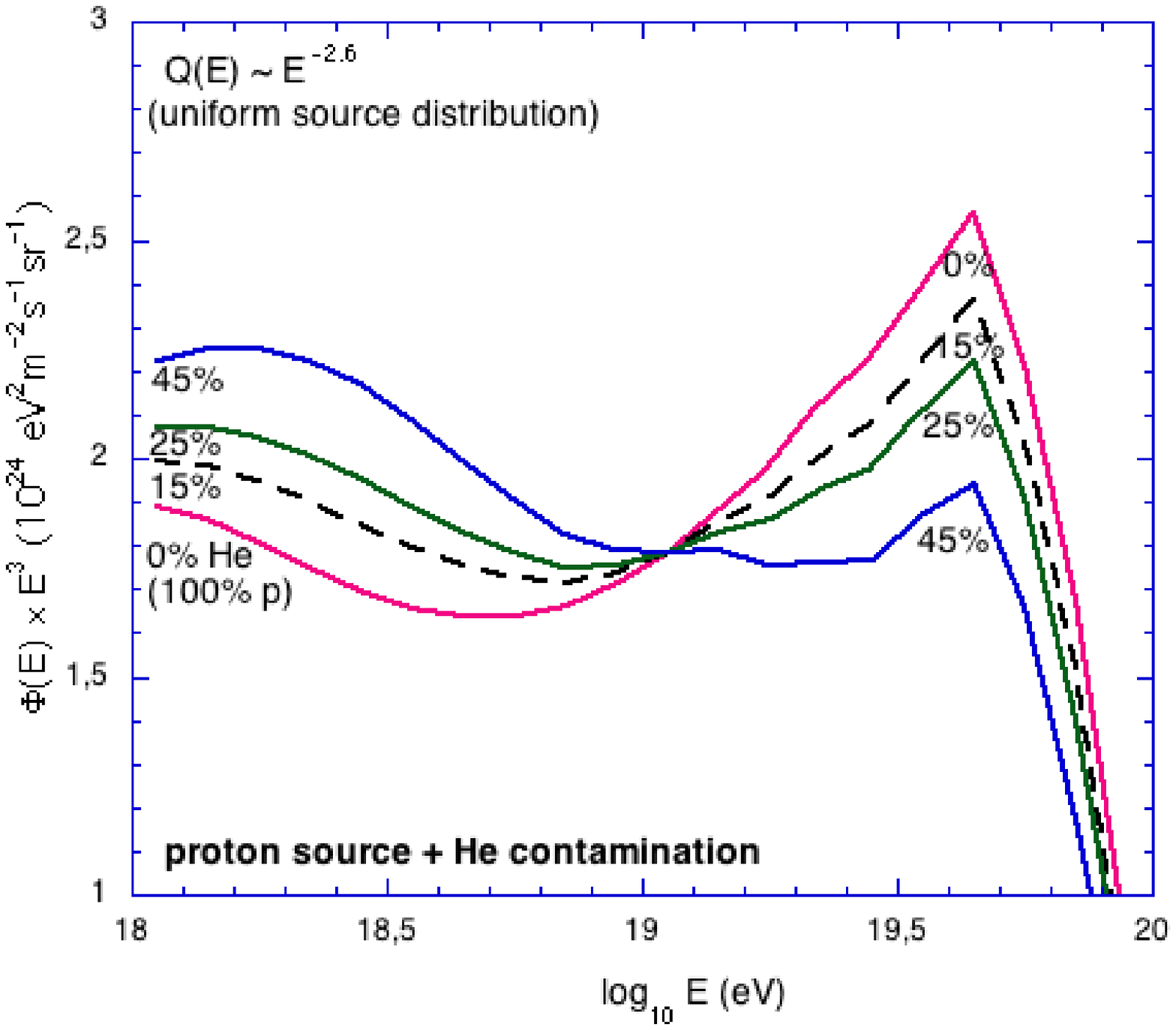}
\caption{Propagated spectra, $E^{3}\, \Phi(E)$, for uniformly distributed sources injecting energetic protons with a spectrum  $\propto E^{-2.6}$ (model B). The different curves show how the shape of the ankle due to the pair-production dip is modified when Fe nuclei (left) or He nuclei (right) are included. The labels give the fraction of Fe (or He) relative to proton at a given energy at the source. All the curves are normalized to HiRes data at $10^{19}$ eV.}
\label{fig:FeHePollution}
\end{figure}

\subsection{Phenomenological limits of model B}

Finally, we turn to the limits of model B, i.e., the conditions under which the ankle can be interpreted as a pair-production dip. The main motivation for this interpretation is the similarity of the shapes between the predicted dip and the observed ankle. This similarity holds as long as the  EGCR component is made of protons only at the source, rather than a mixture of ambient material, i.e., either primordial gas or some gas enriched by standard galactic chemical evolution.
 Therefore, it is interesting to study how the shape of the pair production dip evolves when one includes  heavier nuclei at the source.

In Fig.~\ref{fig:FeHePollution}a, we show propagated spectra obtained for model B (pure proton sources and $x = 2.6$) and an additional component of Fe with the same injection spectrum. The fraction of Fe indicated on the plot is simply the relative abundance of iron at the source at a given energy. We considered three cases: no Fe at all, 15\% and 25\% of Fe. As can be seen, increasing the fraction of Fe leads to a weakening of the spectrum curvature in the ankle region. The reason for such a flattening is that the Fe spectral shape in this energy range is mostly affected by adiabatic losses only and therefore keeps its original shape ($\propto E^{-2.6}$). Even for small fractions of Fe, the pair-production dip interpretation does not fit the spectrum \cite{Berezinsky+05}, requiring an additional Galactic component to reproduce the shape of the ankle.

Similarly, we show in Fig.~\ref{fig:FeHePollution}b that the pair production dip and the global shape of the spectrum are rapidly modified when He nuclei are added to the ECGR source composition. He fractions higher than $\sim$ 15\% at fixed energies strongly modify the position and the amplitude of the expected features, which is a relatively low He injection fraction.  
For instance, if $\sim$ 7-8\% He primordial abundance is injected into an accelerator that generates a spectrum $\propto Z\, E^{-2}$, $\sim$ 30\% He fraction would be produced per fixed energy. 
 For a steeper spectrum the fraction would be larger. And if the injected material is from an interstellar medium similar to that of the Milky Way, the fraction of heavier nuclei would be much higher.
 
While it is customary to think of He nuclei as representing roughly 10\% of the CRs at low energies (say, around 10~GeV/n), this corresponds to the relative normalization of the differential spectra in \emph{energy-per-nucleons}. If one considers the spectra of the different elements at a given \emph{energy}, as we do here (and as it is usually done at and above the knee), the actual ``He fraction'' is much larger. For instance, at energies below the proton knee, the fraction of Helium estimated by combining observations  of a number of detectors and a rigidity dependent knee ranges from $\sim$ 30\% to 50\% \cite{Hoerandel03}. 
 Kascade observations \cite{Kascade}, for any choice of hadronic models used to interpret the data, give  larger He than  proton abundances at the knee ($\sim 1 - 5 \times 10^{15}$~eV), and even larger above the knee, since the proton flux breaks earlier than the He component. 
A fraction of 15\% of He can be considered unexpectedly small for EGCR sources, even if the accelerated material is not enriched by stellar or explosive nucleosynthesis, such as primordial gas,  if the source spectrum is steeper than that of GCRs as required for model B. 
A break  in the source spectrum at energies below $10^{18}$~eV does not alleviate the difficulties, since  the break is likely to be rigidity dependent and to occur at energies two times lower for protons than for He nuclei.

In sum, from the phenomenological point of view, model B requires a significant rejection of nuclei other than protons, including primordial He nuclei. How this may be achieved by the acceleration mechanism remains an open question. In principle, electromagnetic mechanisms do not distinguish between charged nuclei of the same rigidity, so the required discrimination must be found elsewhere, for example, in a strong radiation field around the source and/or a highly magnetized local medium. Here we focus on the different models from the point of view of their predictions for the spectrum and composition features, and leave the problem of possible sources to future studies.

\section{Composition constraints on the EGCR models}
\label{sec:compoConstraints}

\subsection{Composition-related observables: $X_{\max}$ and $\langle \ln A\rangle$}

We now turn to the study of the high-energy CR composition and the evolution of the composition-related observables with energy, as predicted by models A and B as compared to the available data. There are two main composition observables. The first one is the so-called ``$X_{\max}$'', defined as the grammage, in g/cm$^{2}$, measured along the shower development axis (from space to the ground), at which the maximum development of the shower is reached, i.e., the position where the number of secondary particles produced in the electromagnetic cascade is largest. $X_{\max}$ can be measured for each shower by experiments detecting the emission of fluorescence light in the wake of the shower (due to the recombination of the ionized air), such as Fly's Eye, HiRes, and Auger. Showers induced by high-energy particles can develop for a longer time in the atmosphere before fading away, and thus have larger $X_{\max}$ values than showers induced by lower-energy CRs. Conversely, heavy nuclei have a smaller interaction length than protons in the atmosphere and thus produce showers that develop earlier (on average), resulting in smaller values of $X_{\max}$. While the predicted values of $X_{\max}$ for a given nucleus at a given energy is model dependent, their variation with the mass and energy of the incident particle provides an useful observable to constrain the evolution of the high-energy CR composition with energy. As defined above, the evolution of $X_{\max}$ with energy is called the elongation rate.

The second observable is accessible to high-energy CR detectors that measure the distribution of the particles on the ground, including muons, such as Akeno/AGASA. It is based on the estimation of the muon content of the showers, which is expected to be larger for heavier progenitors, due to the larger contribution of the hadronic cascade relative to the electromagnetic one. While a clear identification of the energetic nuclei is quite hard to obtain with such a technique, one may estimate the average mass of the incident CRs (or its logarithm, $\langle \ln A\rangle$), and measure its variation as a function of energy.

Below we compare the predictions obtained from models A and B for these two observables with the available data from HiRes and Akeno/AGASA (for an  exhaustive description of  current experimental results, see \cite{Dova05,Watson}).  The two observables we chose to compared the propagated spectra of each model with have the best published data in the relevant energy range. There are other composition related observables that should be studied in the future as higher statistics are accumulated such as the width of $X_{\max}$ as a function of energy. In addition, 
extensive air shower detectors using water Cerenkov tanks (Haverah Park \cite{HaverahPark,Ave2003b} and Auger \cite{PAO}) have access to other types composition related observables, such as the rise time of the signals, the steepness of the lateral distribution function, and the curvature of the shower front, which appear to correlate with $X_{\max}$ and/or muons richness of the showers (see \cite{NaganoWatson}, for a review of the experimental techniques).

\subsection{Cosmic-ray composition after propagation}

\begin{figure}[t]
\centering
\includegraphics[width=0.48\linewidth]{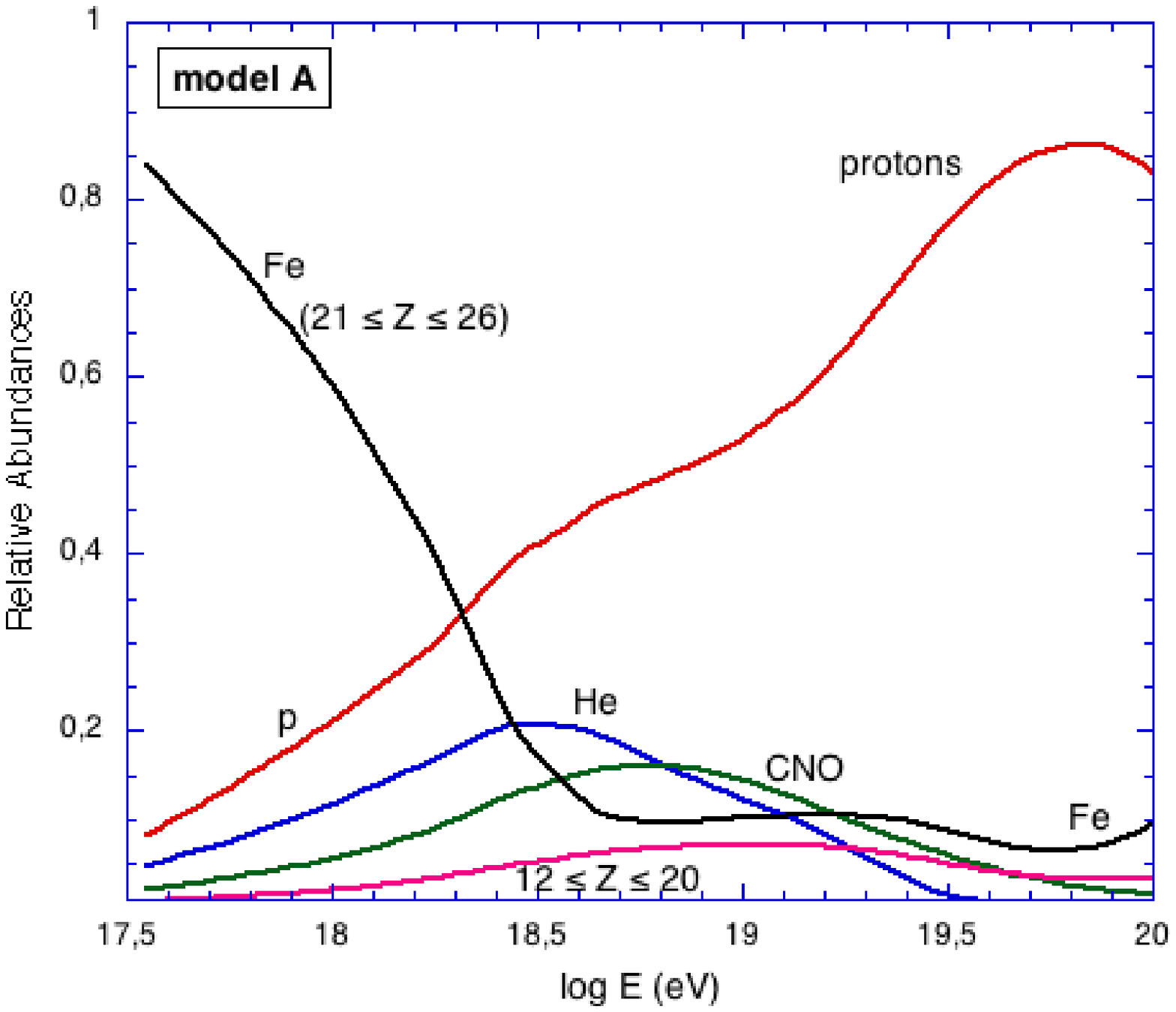} \hfill
\includegraphics[width=0.48\linewidth]{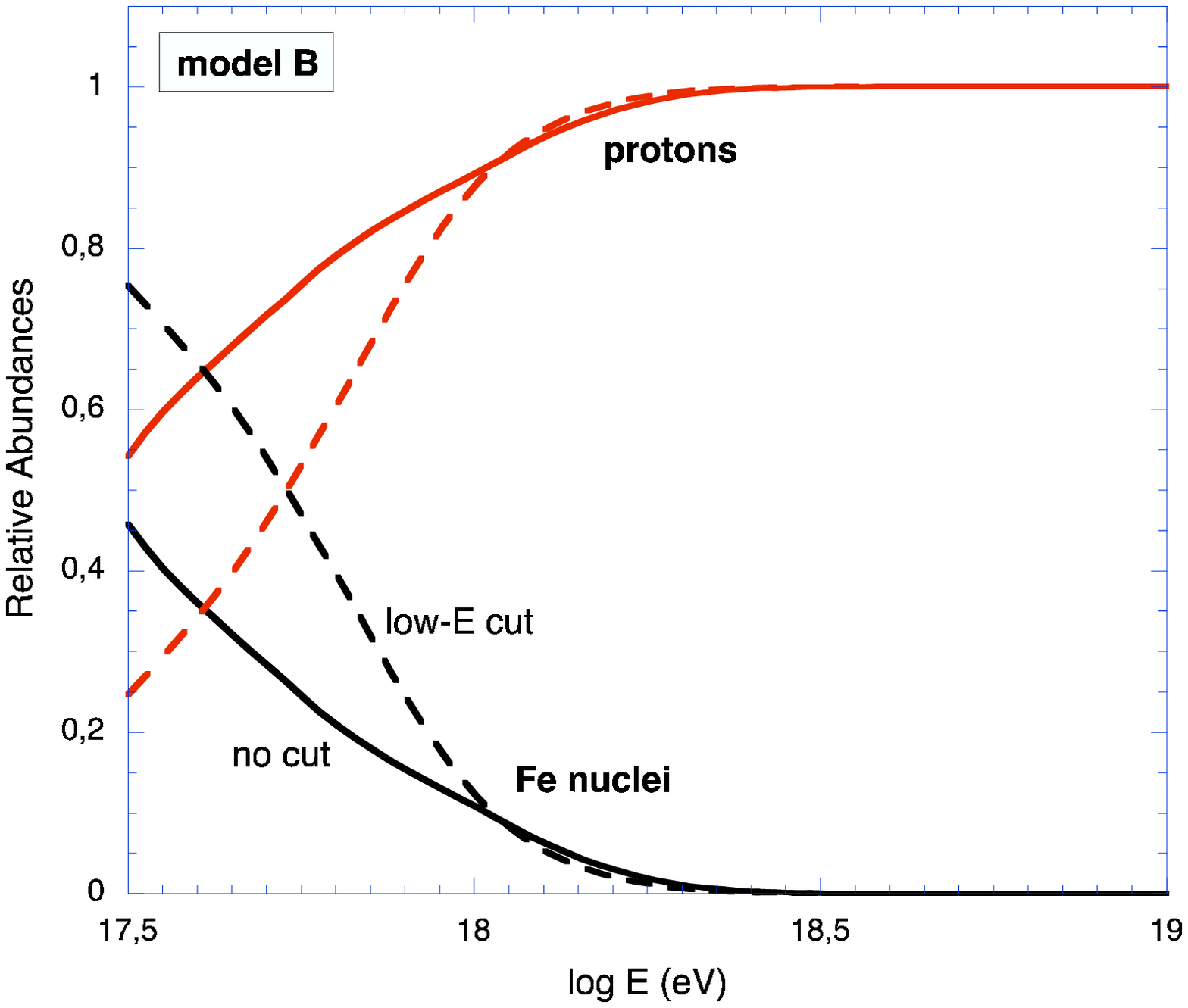}
\caption{Relative abundances of high-energy cosmic rays observed on Earth taking into account propagation effects. Left: for model A (where  the indicated ``Fe component'' includes all nuclei with $21 \leq Z \leq 26$). Right: for model B, with a single $x = 2.6$ power-law spectrum above $10^{17}$~eV (solid line) or with a ``low-energy cut'' with  $x = 2.0$ below $10^{18}$~eV (fit to the HiRes spectrum).}
\label{fig:compo}
\end{figure}

In Fig.~\ref{fig:compo}, we show the composition of the propagated CRs above $3\,10^{17}$~eV, as predicted by models A and B. We assumed that the GCR component above $3\,10^{17}$~eV is made of Fe nuclei only, in agreement with experimental results indicating a progressive transition from light to heavy nuclei at the knee. Our results not sensitive to this assumption as long as  the Galactic component in this energy range is heavier than H or He. A slightly lighter (or even heavier than pure Fe \cite{Hoerandel05}) GCR composition would lead to essentially the same predictions for the observables analysed here.

The case of model A is shown in Fig.~\ref{fig:compo}a, where one can see the relatively slow decrease of the Fe fraction and the even slower increase of the proton fraction as the energy increases. The contribution of intermediate nuclei is highest around $3\,10^{18}$~eV,  precisely \emph{at the ankle}, where they jointly represent more than 30\% of the cosmic-rays for this particular composition model. By contrast, model B implies a much quicker transition from an Fe-dominated composition to a pure proton composition, as shown in Fig.~\ref{fig:compo}b. Above $\sim 1.5 \,10^{18}$~eV, protons make up 100\% of the observed cosmic-rays, instead of $\sim 40$\% in the case of model~A. 

Another important difference between the two models is the composition at $\sim 1- 3\times 10^{17}$ eV. Model A predicts a very large contribution of Fe up to $3\,10^{17}$~eV, and more than 50\% of Fe around $10^{18}$~eV, while model B predicts a larger contribution of protons down to $3\,10^{17}$~eV, and a significant proton fraction at $10^{17}$~eV, which is disvafored by the observations. This point depends slightly on the assumed source evolution as strong evolution can flatten the required injection  spectrum. 

However, model B requires a low-energy cut or hardening of the EGCR spectrum which tends to decrease the role of EGCRs with respect to the heavier GCRs between $10^{17}$ and $10^{18}$~eV. This variation of model B is shown by the dashed lines in Fig.~\ref{fig:compo}b, which corresponds to a source spectrum with a break at $10^{18}$~eV and a spectral index of $x = 2.0$ below this energy. A similar decrease of the EGCR component below $10^{18}$~eV may be due to  extragalactic magnetic fields that prevent low-energy CRs to reach our Galaxy from EG sources. 
If the Kascade data are confirmed, the problem of the very light composition predicted by model B at $3\,10^{17}$~eV, may be overcome either by a change of the spectrum at the source or by invoking plausible propagation effects. However, these changes imply an even sharper transition from a pure-Fe to a pure-proton component, which leads to distinctively different predictions, as compared with model A, for the composition-related observables in the ankle region. These predictions are compared with the available data below.

\subsection{Position of the shower maximum: $X_{\max}$}

\subsubsection{Model predictions}

To compute the expected value of $\langle X_{\max}\rangle$ as a function of energy corresponding to the different models, we first simulated a set of 50,000 extensive air showers with the shower development codes Corsika \cite{Corsika}, Aires \cite{Aires} and CONEX \cite{Conex1}, for H, He, CNO, Si and Fe primaries. We used  three different models for the high-energy hadronic interactions, namely QGSJet-01 \cite{QGSJet}, Sibyll 2.1 \cite{Sibyll}, and the recently developed QGSJet-II \cite{QGSJetII}. Combining the corresponding values of $X_{\max}$ obtained for each type of incident nucleus at each energy, we calculated the average $X_{\max}$ weighted by the relative abundance of the different nuclei, as predicted by the models (cf. Fig.~\ref{fig:compo}).

\begin{figure}[t]
\centering
\includegraphics[width=0.48\linewidth]{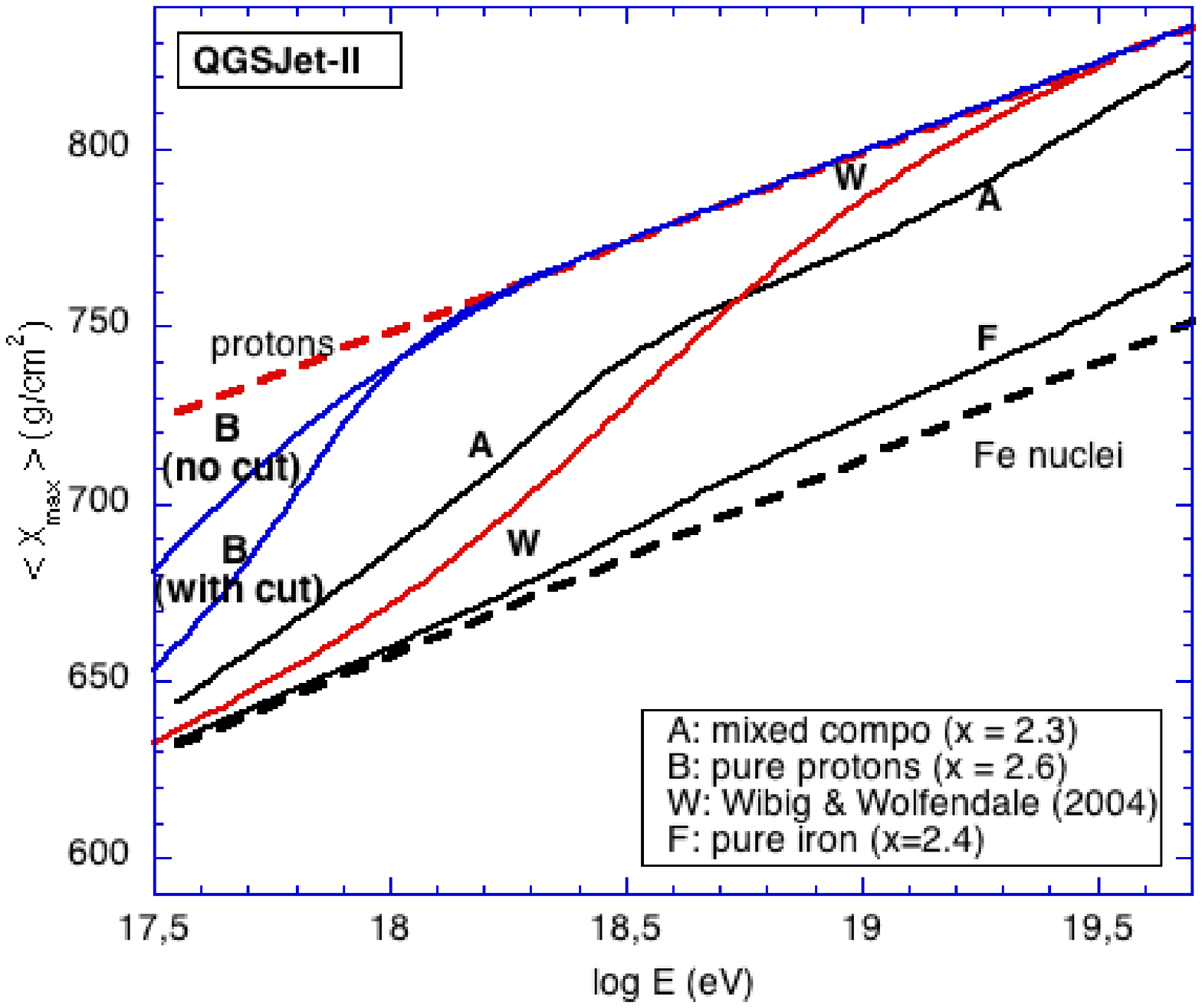} \hfill
\includegraphics[width=0.48\linewidth]{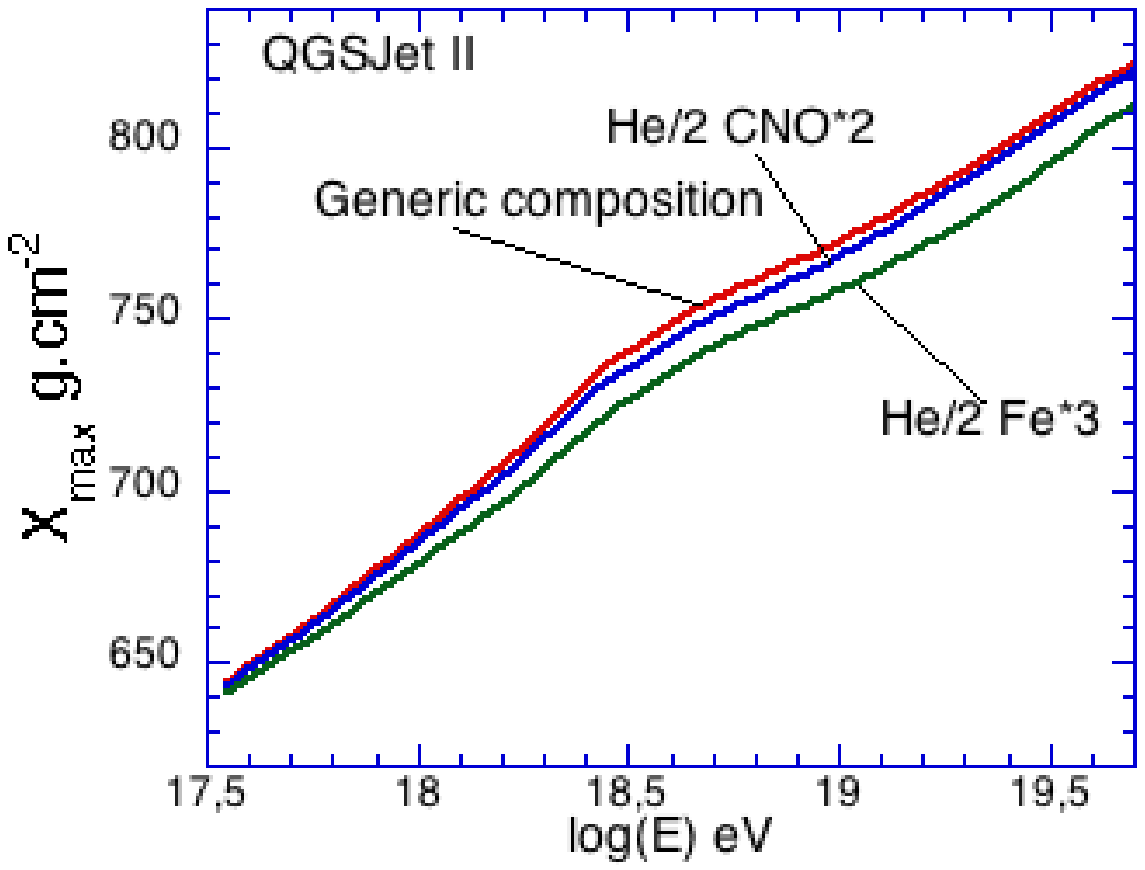}
\caption{Left: depth of the shower maximum, $X_{\max}$, in g/cm$^{2}$, as a function of energy for model A (labeled A), model B (labeled B),   the model of Wibig and Wolfendale \cite{WW} (labeled W) and a pure iron source composition. The upper and lower dashed lines correspond to the mean shower maximum of incident protons and Fe nuclei, respectively, with  $X_{\max}$  from Corsika showers and QGSJet-II  hadronic interaction model. Right: comparison of model A predictions for three different source compositions, as indicated.}
\label{fig:XMax}
\end{figure}

The results are shown in Fig.~\ref{fig:XMax}a for  Corsika showers with hadronic model QGSJet-II. The dashed lines give the values of $X_{\max}$ for proton- and Fe-induced showers as a function of energy. As expected, model B predicts a rather steep evolution of $X_{\max}$ between $10^{17.5}$ and $10^{18}$~eV, as a result of the sharp decrease of the Fe abundance through the transition. A version of model B where the low-energy part of the spectrum is reduced (either at the source or by magnetic processes) would lead to an even steeper curve (with an elongation rate $\sim160$ for the QGSJet and Sibyll models between $10^{17.5}$ and $10^{18}$ eV). 
With model A, the transition from $X_{\max}$(Fe) to $X_{\max}$(protons) is smooth, over two orders of magnitude in energy. A characteristic feature of this model is the existence of an inflection point in the $X_{\max}(E)$ curve, at an energy close to $4\,10^{18}$~eV, which is roughly at the ankle minimum. The inflection  corresponds to a ``delay'' in the lightening of the high-energy CR composition with energy  due to the appearance of the He and CNO components, as seen in Fig.~\ref{fig:compo}a. In the transition region, the $X_{\max}$ evolution with energy is significantly gentler than in the case of model B, since the composition does not immediately turn to pure protons. A flattening is visible between $10^{18.4}$ and $10^{19}$~eV,  when the observed CRs are dominated by the extra-galactic component and the composition does not change significantly (mixed composition regime). Finally, the $X_{\max}$ evolution gets steeper again as the composition gets lighter, due to the suppression of heavier components at the onset of their photo-disintegration processes with the CMB (beginning with He nuclei). 

As shown in Fig.~\ref{fig:spectraMixedCompo}a the precise source composition assumed for model A does not have a significant impact on the shape of the propagated spectrum and on the ankle interpretation. In Fig.~\ref{fig:XMax}b, we further show that the predicted evolution of $X_{\max}$ with energy is also largely insensitive to the EGCR composition details, by comparing the predictions for the three different compositions discussed above.  In all three cases, the ``delay in lightening'' feature is present at the same energy, with essentially similar amplitude. Note, however,  that the signature of the delay in lightening would be less pronounced in the case of a originally light mixed composition. In this case, a noticeable flattening is expected at $E_{ankle}$ but the lightening above $10^{19}$ eV would be more subtle. Conversely, if the relative contribution of secondary protons is low above $\sim3\,10^{18}$ eV (e.g., if the maximum energy reached by nuclei heavier than proton is low), the relative abundance of protons in the EGCR flux between $10^{18}$ and $10^{19}$ eV could decrease due to the effect of  pair production interactions. In this case, the ``s" shape in the evolution of $X_{\max}$ would be more pronounced than for the cases we display here and a slightly higher proton fraction at the sources would be required to fit acurately the UHECR spectrum. 

\begin{figure}[t]
\centering
\includegraphics[width=0.48\linewidth]{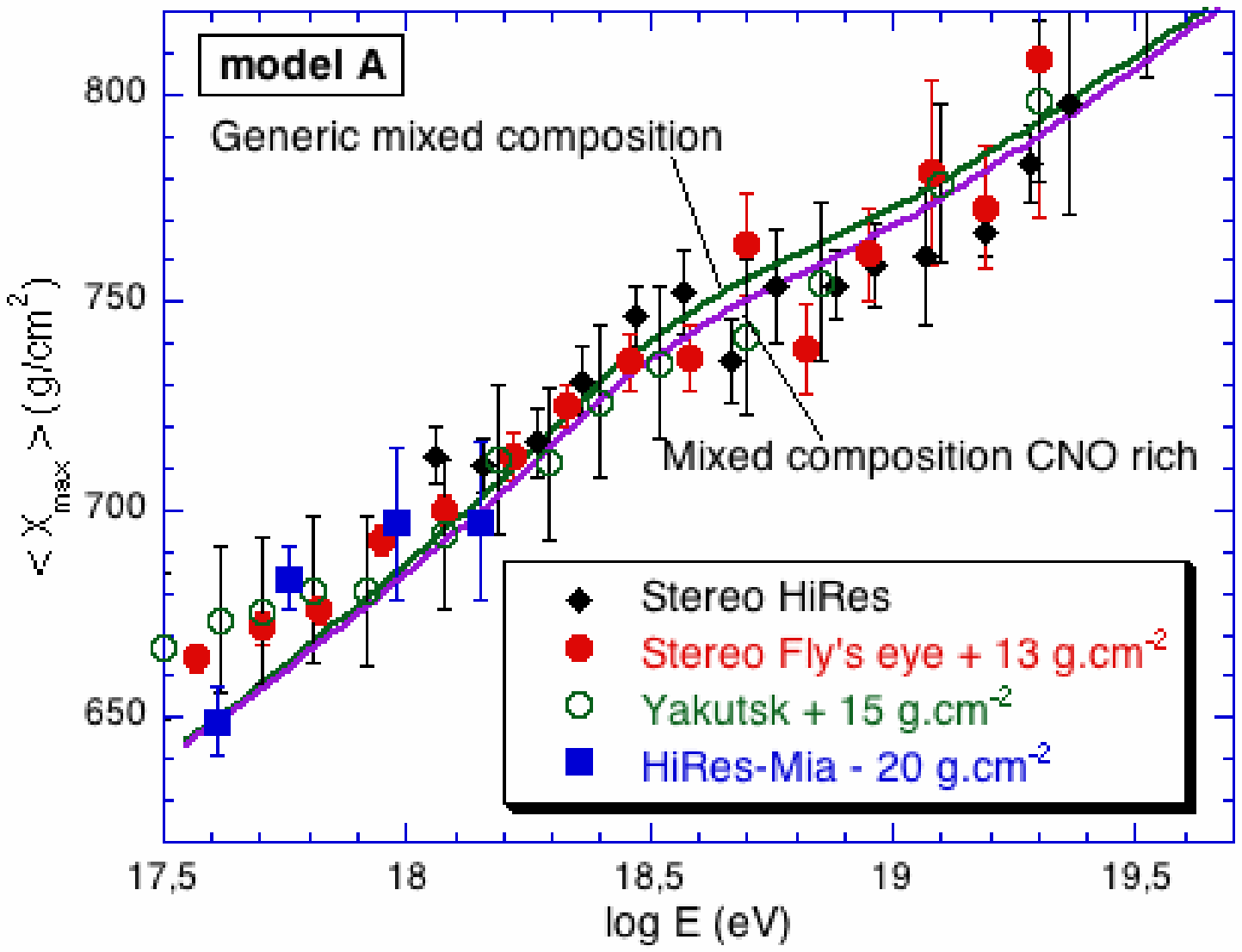} \hfill
\includegraphics[width=0.48\linewidth]{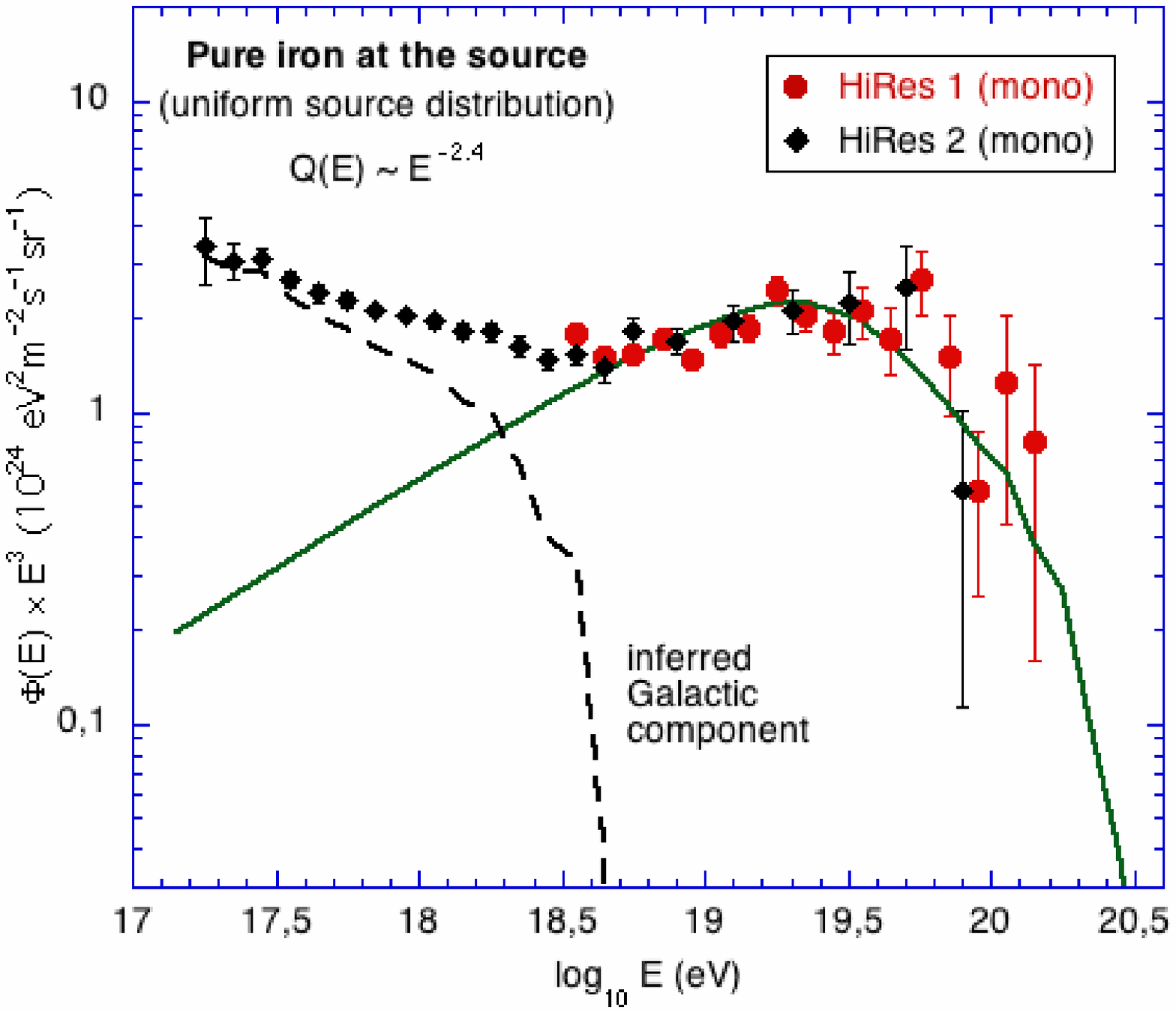}
\caption{Comparison of the predicted $X_{\max}$ evolution  with data from Fly's Eyes, HiRes, HiRes-Mia, and Yakutsk slightly rescaled (see legend) compared with the predictions of Model A (generic and CNO rich composition, QGSJet-II).  Right: Propagated spectra, $\Phi(E) \, E^{3}$, for model F (pure iron EGCR source composition), compared with the HiRes data the dashed line corresponds to the inferred GCR component,}
\label{fig:XMax_last}
\end{figure}

In Fig.~\ref{fig:XMax}a, we also show the case of a pure iron source, as an illustration. It indeed appears that a model in which the EGCR sources only accelerate Fe nuclei can also explain the observed CR spectrum at high energy, in a way rather similar to our model A, i.e., with an interpretation of the ankle as the GCR/EGCR transition and a somewhat steeper source spectrum, $\propto E^{-2.4}$ (although the fit of the HiRes spectrum is worse at the highest energies). This corresponding propagated EGCR spectrum and the associated GCR component are shown in Fig.~\ref{fig:XMax_last}b. While such a model appears viable from the point of view of the spectrum, it actually contradicts the data on the evolution of $X_{\max}$ and $\langle\ln A\rangle$ with energy (see below) as it does not exhibit a clear evolution toward  light elements in the transition region.  Indeed, the contribution of the secondary protons below $10^{19}$ eV is low ($\sim 7-8 \%$) and the corresponding evolution of $X_{max}$ in the transition region is very slow. This disagreement can be partially overcome if a strong source evolution and a harder spectral index (2.0) are  assumed. In this case, the large number of secondary nucleons produced ($\sim30\%$) make the composition significantly lighter and more compatible with the data (also this modified model would still be disfavored compared to the other models we consider). Finally It is important to keep in mind that the predictions for a pure iron source composition would also be more sensitive to EGMF  than the other models we consider here.

Another curve is shown in Fig.~\ref{fig:XMax}a with the label ``W''. It corresponds to the model of Wibig and Wolfendale \cite{WW}, where EGCRs are essentially protons, but with a harder source spectrum than model B, so that the ankle is still interpreted as the GCR/EGCR transition. In this model, the galactic component (essentially Fe nuclei in the range of interest here) keeps a non-negligible contribution to the CR flux up to a high energy, $\sim 3\,10^{19}$~eV. Given the current precision of $X_{\max}$ measurements, the predictions obtained with this model are not very different from those of model A. The  characteristic change of slope or delay in lightening between $3\,10^{18}$ and $10^{19}$~eV  is however not present in model W.  Detailed observations around $10^{19}$~eV, with the kind of  statistics that will be available from the Pierre Auger Observatory, should allow one to distinguish between the two models in the future.

\begin{figure}[t]
\centering
\includegraphics[width=0.48\linewidth]{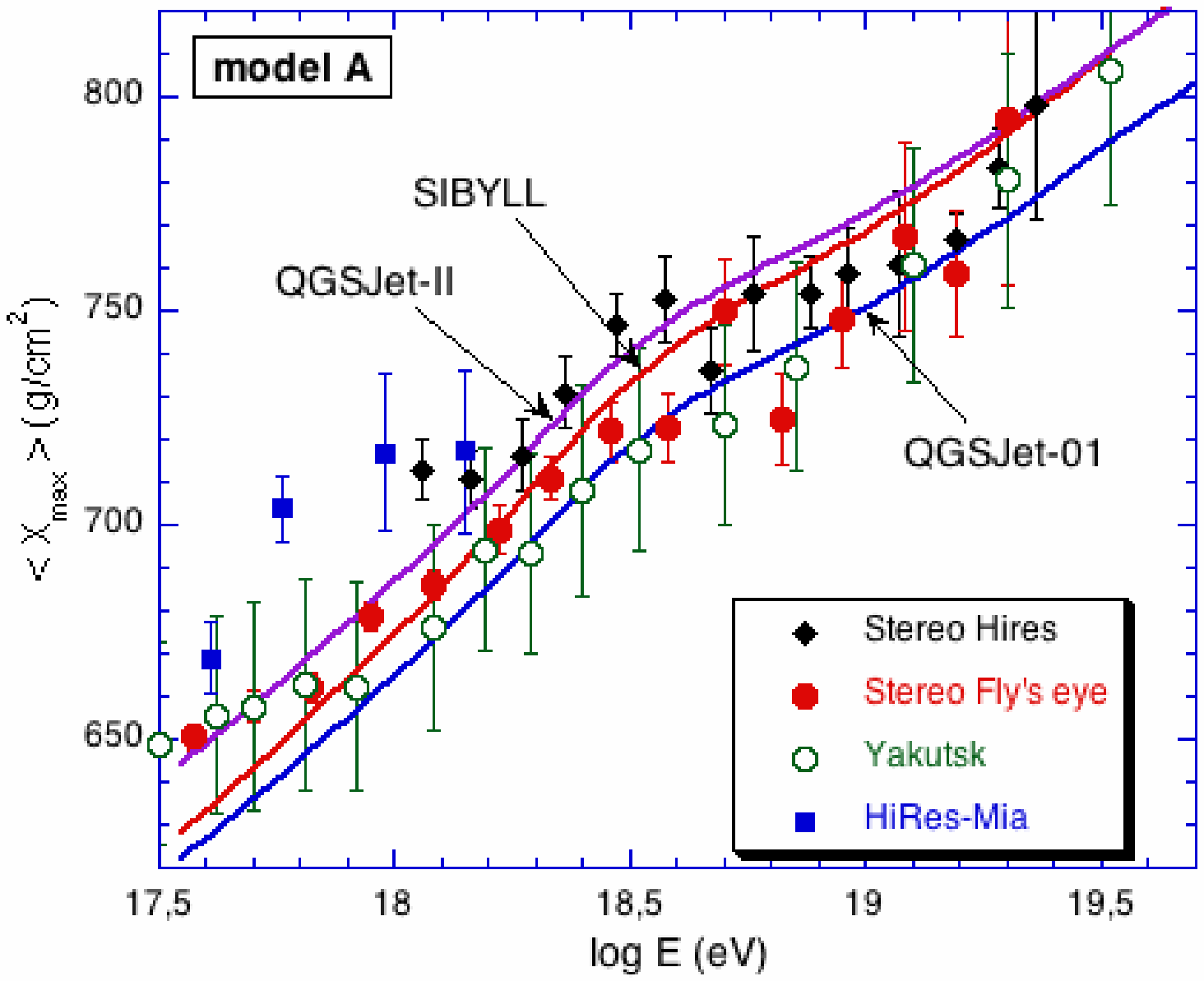} \hfill
\includegraphics[width=0.48\linewidth]{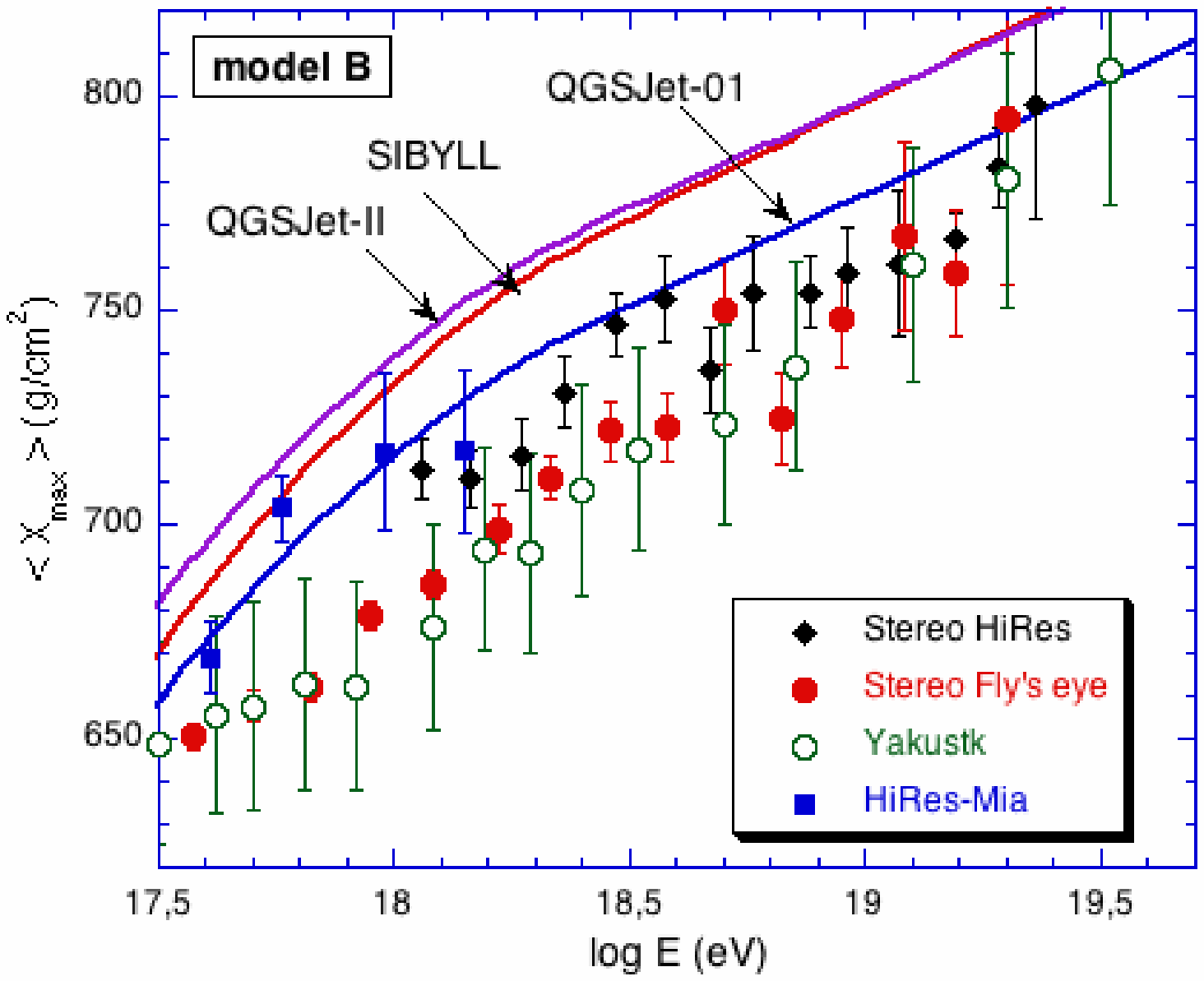}
\caption{Comparison of the predicted $X_{\max}$ evolution with energy with the data from Fly's Eyes, HiRes, and Yakutsk compared with the predictions of the models A (left) and  B (right, without a low energy cut) with three different hadronic models (see figure and text.)}
\label{fig:XMaxComparData}
\end{figure}

\subsubsection{Comparison with the data}

In Fig.~\ref{fig:XMaxComparData}, we compare the  predictions of different models with the available data. In Fig.~\ref{fig:XMaxComparData}a, it can be seen that the predictions of model A with the three hadronic models  agree  remarkably well with the data from Fly's Eye \cite{Bird+93}, Yakutsk \cite{Afanasiev93} and Stereo HiRes  \cite{HiRescomp} from the point of view of  the normalization and the general shape of the evolution (including the ``delay in lightening'' feature). The latter can be further checked provided that mild shifts in the data within the claimed systematics are performed to bring the experiments in agreement (see \ref{fig:XMax_last}a). 
It can be further seen that HiRes Stereo data appear compatible with a flattening of the $X_{\max}$ evolution at $\sim10^{18.5}$ eV  corresponding to the energy of the ankle claimed  by the experiment, which is more pronounced than the mixed composition cases we displayed, and a lightening above $10^{19}$ eV. The agreement between the different experiments is noticeable (see \cite{Sokolsky05}) on the whole energy range and this is all the more true when a shift of 15-20 $g.cm^{-2}$ is applied to HiRes-Mia \cite{HiRes-Mia}. In this case the latter appears quite compatible with the predictions of the mixed composition model (except at $10^{17.7}$ eV) . Furthermore, it is important to note that the elongation rate claimed by this experiment between $10^{17}$ and $10^{18}$ eV (slightly above 90, see \cite{HiRes-Mia}) is also  compatible with the predictions of the model A (typically between 90 and 100). The conclusion that the data are well accounted for  by model A is not affected much  by different hadronic models which give an overall normalization of $X_{\max}$ with differences of the order of 20~g/cm$^{2}$, a range which is well within the measurements uncertainties. 

In the case of model B (without a low energy cut), it can be seen on Fig.~\ref{fig:XMaxComparData}b that predictions are systematically higher than the measurements. However, the HiRes-Mia data at low energies, between $3\,10^{17}$ and $10^{18}$~eV, are well reproduced by model B, as they favour a sharp transition from heavy to light CRs in this range. Note that a low energy break at $10^{18}$ eV would produce too fast a transition  even for the HiRes-Mia data. At higher energies, model B predictions are systematically  high compared with the data, since the EGCRs are made of protons only in this model, while the data seem to indicate a heavier composition. 
A $\sim15$ g/cm$^{2}$ shift upward of the HiRes Stereo data would give a better agreement with the predictions of the QGSJet-01 model. However, a shift of $\sim30\,g.cm^{-2}$ is needed in the case of the QGSJet-II (which the latest version of the QGSJet models) and SIBYLL models. In this case, shifts of $\sim40\,$g/cm$^{2}$ for Fly's eye and $\sim45\,g.cm^{-2}$ for Yakutsk would be necessary and would not modify the complete absence of evidence for a very steep transition completed at $\sim10^{18}$ eV in the data of the two latter experiments.  

As mentioned above the absolute normalization of $X_{\max}$ is quite uncertain and model-dependent. Overall shifts in either the data or the model predictions are both allowable in principle. However, the so-called \emph{elongation rate}, is relatively robust. Because of the sharp transition from a heavy GCR component to a light EGCR component, model B predicts a very large elongation rate ($\sim120-130$ without low energy cut, $\sim160$ for $E_{break}=10^{18}$ eV) at below $10^{18}$~eV, where the transition has to be completed and the elongation rate then settles at to the much lower, pure-proton value. Higher precision measurements and a better control on the hadronic model at high-energy will be needed to assess the compatibility of the different models with the data and to solve the significant disagreement between the HiRes-Mia data points around $10^{17.7}$~eV and the measurements of the other experiments. However, taken at face value with the most recent versions of the popular hadronic models, the current data on $X_{\max}(E)$ favour model A and a smooth transition from GCRs to EGCRs associated with the ankle. We finally claim that the general \emph{shape} of the $X_{\max}$ evolution with energy is a powerful observable to distinguish between the models\footnote{It should be pointed out that $X_{\max}$ distributions and their associated widths, that are known to be quite hadronic model independent, would be poorly efficient to distinguish a pure proton composition from a proton dominated mixed composition, especially if low and intermediate mass nuclei significantly contribute.} as they all predict specific behaviour and features. These features in the slope of the evolution of $X_{\max}$ should be accessible to high statistics experiments, such as the Pierre Auger Observatory \cite{PAO}.

\subsection{Muon content of the showers and typical CR mass: $\langle \ln A\rangle$}

As mentioned above, ground array detectors are also able to constrain the UHECR composition. In the following we will only compare our predictions with the Akeno-AGASA results
but an exhaustive study of the current experimental results can be found in \cite{Dova05,Watson}. The composition analysis of Akeno/AGASA relies on the measurement of the number of muons, from which evolution with energy one can infer the evolution of the proton and iron fractions ($x_{H}$ and $x_{Fe}$, respectively) among high-energy CRs, using shower and detector simulations. From this estimate, one can then compute $\langle \ln A\rangle$ by noting that $\langle \ln A\rangle=x_{Fe}\times \ln(56)$. On the other hand, the above-mentioned models can be used to predict the evolution of $\langle \ln A\rangle$ for a both model A (mixed composition) and model B (pure proton sources), with or without a low-energy break.

\begin{figure}[t]
\centering
\includegraphics[width=0.48\linewidth]{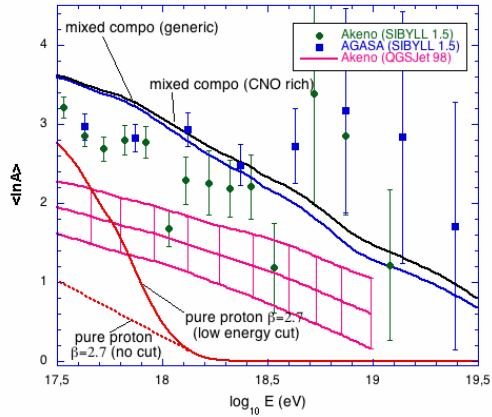} \hfill
\caption{Evolution of the mean logarithmic mass, $\langle\ln A\rangle$, with energy. The Akeno and AGASA data (courtesy of A.~G. Mariazzi) are interpreted using two hadronic models: Sibyll-1.5 (data points) or QGSJet-98 (shaded area), as indicated. The lines correspond to the prediction for models A and B (with or without a low-energy break in $E^{-2}$ below $10^{18}$~eV.}
\label{fig:muonData}
\end{figure}

In the case of model B, the comparison with the proton and iron fractions deduce from Akeno and AGASA is straightforward, since there are indeed only (extragalactic) protons and (galactic) Fe nuclei among the cosmic-rays (see above). In the case of model A, the comparison is made problematic in principle, for observational reasons. Although it is straightforward to deduce the real proton and iron fractions from our propagation code, it is hard to determine which fraction of the low and intermediate mass nuclei ($\sim 40\%$ of the composition at the sources) would have been counted as proton or iron in the experimental analysis. To do so requires an accurate simulation of the extensive air showers and of the Akeno and AGASA detectors, which we cannot address here. In any case, the computation of proton and iron fractions from the muon content of the shower particles depends of hadronic models. For the comparison of the experimental fractions with our predictions, we use either QGSJet-98 \cite{Shinozaki} or Sibyll-1.5 \cite{Dawson98}. As can be seen on Fig.~\ref{fig:muonData}, the inferred fractions of iron are indeed very much model-dependent, with differences between the two models reaching more than $50\%$. As mentioned in \cite{Berezinsky+05b}, the composition appears proton rich at $10^{17.5}$ when the QGSJet-98 model is used. However, even if the fractions themselves are model-dependent, the general shape of the energy evolution of $\langle\ln A\rangle$ is similar in both cases (although slightly steeper with the Sibyll model), with a smooth evolution from $10^{17.5}$ to $10^{19}$~eV and a significant fraction of nuclei above $10^{18}$~eV. This contrasts with the sharp drop of the heavy fraction expected from model~B (with or without the low-energy break). This difference does not seem to be easy to explain simply by a possible bias pertaining to the poor shower and detector simulations. It is worth mentioning that the Sibyll-1.5 hadronic model generally leads to $\sim 44\%$ less muons than QGSJet-98, and $\sim 17\%$ less than Sibyll 2.1 \cite{Alvarez+02}, which is very similar to QGSJet-II. One can then deduce that the current versions of Sibyll and QGSJet would lead to lower iron fractions than obtained with Sibyll-1.5, but still well above those obtained with QGSJet-98.

On the other hand, model A appears to predict a much smoother evolution of $\langle \ln A\rangle$ with energy, in better agreement with the available data. For the reason mentioned above, it is however difficult to directly compare the predicted curve with the values inferred from the experimental results (given the significant number of intermediate-mass nuclei). One may just note that they appear compatible. The Akeno/AGASA data also favour a smooth transition over the energy range of interest, which would also be compatible with the above-mentioned model by Wibig and Wolfendale \cite{WW}.For the sake of completeness, we finally note that a recent muon analysis from Yakutsk indicates a composition getting lighter between $10^{17}$ and $10^{18}$ eV and dominated by light nuclei above $3\,10^{18}$ eV \cite{YakutskMuons}, which is predicted by both models. The Haverah Park composition studies using rise times and LDF steepness  \cite{HaverahPark,Ave2003b}  disfavour Model B's  sharp transition between $10^{17}$ and $10^{18}$~eV.

\section{Summary and discussion}

In this paper, we have studied the phenomenology of the GCR/EGCR transition within two types of models for the extragalactic component: model A assumes a mixed source composition, with nuclear abundances similar to that of the GCRs, while model B assumes that the EG sources accelerate only protons. From the point of view of the CR energy spectrum, both models appear to be able to reproduce the high-energy CR data equally well, although with a different source spectrum and a correspondingly different interpretation of the ankle. Within model A, the source spectrum is typically in $E^{-2.3}$, and the ankle marks the transition from GCRs to EGCRs. Within model B, the source spectrum is steeper, in $E^{-2.6}$ or $E^{-2.7}$, and the ankle is fitted all the way down to $10^{18}$~eV by the EGCR component alone, exhibiting a characteristic dip due to the pair production interactions of ultra-high-energy protons with the cosmological microwave background.

Although we have assumed that the intensity of the sources did not evolve in the last few Gyr, none of the results presented here would be significantly modified if the sources actually evolved with redshift, either in number or in intensity. The exact slope of the required source spectrum could be slightly modified, but model A would also require a significantly harder spectrum, i.e. less luminous sources (from the low-energy extrapolation). We also studied the phenomenological limits of the models, and showed that the main features of model A were quite robust to a reasonable modification of the source composition, while the pair-production dip interpretation of the ankle provided by model B would be destroyed if a relatively small fraction of He or heavier nuclei were also accelerated in (and escaped from) the sources.

Since the available data on the CR energy spectrum are not accurate enough yet to discriminate between models A and B, we focused on the constraints that may be derived from composition estimates. We considered two complementary composition-related observables, $X_{\max}$ and $\langle \ln A\rangle$, accessible respectively to fluorescence detectors (e.g., HiRes, Auger) or ground arrays (e.g., AGASA, Auger). We obtained very different predictions for the evolution of these observables with energy, mostly because of the different scenarios implied for the transition from a heavy GCR component to EGCRs  : model B requires that the transition towards a light (pure proton) EGCR component be completed at an energy as low as $10^{18}$ eV, whereas the heavy GCR component extends up to $3-6\,10^{18}$ eV in model A. The early GCR/EGCR transition of model B also implies an even steeper final drop of the Galactic component if a break of the source spectrum occurs at an energy larger than $10^{17}$~eV (remind that such a break is required for energetics reasons and possibly also not to violate the proton fraction measured at $10^{17}$~eV).

In any case, model B predicts a dramatic change of composition between $10^{17}$ and $10^{18}$~eV, which we have shown in Sect.~\ref{sec:compoConstraints} to be disfavored by composition analysis of  most experiments expect for the very high HiRes-Mia elongation rate between $10^{17.5}$ and $10^{18}$ eV (when no low energy cut is applied). By contrast, the smooth transition implied by the mixed EGCR source composition of model A appears more in conformity with most currently available data.

It should be acknowledged, however, that the composition measurement remain difficult at high energy, and in particular the absolute normalisation of the $X_{\max}$ observable is hadronic-model dependent. While the predictions of model A are in excellent quantitative agreement with the current data, an important result of our study is that the \emph{shape} of the energy evolution of $X_{\max}$ is also very different between both models. A distinctive prediction of model A is the presence of an ``s'' feature in the $X_{\max}$ curve, corresponding to ``delay in lightening'' of the high-energy CRs, directly related to the heavy nuclei accelerated at the sources. This feature may be tested by future, higher-statistics experiments such as Auger, (to some extent) independently of the assumed hadronic interaction model.

The understanding of the origin of the ankle is very important for the phenomenology of not only the ultra-high-energy CRs, but also the Galactic ones. We wish to stress here that it cannot be obtained by the study of the sole spectrum, but benefits a lot from composition studies. It was the purpose of this paper to derive general predictions within the framework of two important and recently discussed models, independently of any consideration about their plausibility or the actual sources and acceleration mechanism involved. We argued elsewhere that model A is from our point of view more natural than model B, because i) a mixed-composition source is expected for processes accelerating particles out of the interstellar medium, ii) a source spectrum in $E^{-2.3}$ seems easier to interpret (notably in terms of relativistic shock acceleration) than a steeper spectrum in $E^{-2.6}$, which would also lead to an energetics problem if not cut at low-energy, and iii) an ankle-like feature is \emph{a priori} natural for a transition between two CR components. However, we should not underestimate the argument that the ankle is very well reproduced by a single component of extragalactic protons over its whole energy range, with a very limited number of free parameters \cite{Berezinsky+02,Berezinsky+04}. We nevertheless note here that the pure proton model meets some difficulties to account for the available composition measurements. It is a fact, however, that the experimental data are still quite uncertain and do not lead to a completely coherent picture above $10^{18}$~eV, even though they all favour a significant fraction of non-proton nuclei at the ankle and a smooth GCR.EGCR transition. Future experiments with larger statistics and higher resolution will be needed to fully understand the origin of the ankle: Kascade-Grande \cite{Grande} should be able to measure the spectrum and constrain the composition above $10^{17}$~eV, where the transition is expected to begin for the two models; and the Pierre Auger Observatory will be able to measure the elongation rate above $5\,10^{17}$~eV using the hybrid detection technique with a very high statistics above $10^{19}$~eV. The joint use of the fluorescence and ground array techniques will also be crucial to constrain the hadronic models, which are key to detailed understanding of the high-energy CR composition. Furthermore, the very high expected statistics coupled with the full sky coverage provided by the two sites of the Auger (North and South) should reveal the sources of UHECRs and thus provide the last ingredient necessary to the understanding of their origin.

Finally, we briefly comment on the possible influence of extragalactic magnetic fields (EGMF), which have been neglected here. Definite predictions cannot be proposed, mostly because we still lack strong constraints from both observational and theoretical studies about the strength, the coherence length and the structure of the EGMF (see for instance the large differences obtained for similar approches of magnetic fields in large scale structures: \cite{Sigl05,Dolag05}). Qualitatively, however, we noted above that a magnetic horizon effect is expected at low energy (\cite{Lemoine+05,Parizot04}), which could affect the extragalactic spectrum and make the GCR/EGCR transition sharper. This would thus amplify the specific features of model B described above, and possibly make it even more difficult for it to account for the composition-related data. Likewise, in the case of model A, the magnetic horizon would be rigidity-dependent and thus show at higher energy for heavier nuclei. The mixed composition would thus get heavier with increasing energy (with the heaviest nuclei entering the horizon at higher energy), which could partly (depending on the intensity of the magnetic field) counterbalance the steepening of the $X_{\max}$ evolution curve discussed above in the case of a pure proton composition, and also make the characteristic ``s'' shape of model A more pronounced if heavy nuclei start becoming abundant in the mixed composition regime (corresponding to the flattening of the $X_{\max}$ evolution; see above). In any case, whatever the effect of the hypothetic magnetic horizon, it would not lead to a confusion between the predictions of models A and B since the energy scale of the transition would still remain different. Furthermore the expected $X_{\max}$ features (steep-flatter-steep) should remain present unless nuclei enter the horizon at energies above their photo-disintegration threshold, which would require quite a large EGMF (larger than used in \cite{Lemoine+05}). A large EGMF could also increase the path length of the UHE nuclei at high energy, as studied in \cite{Gunter} and thus shift the photo-disintegration cut-off of the different types of nuclei to a lower energy. However, as we have shown above, the predicted spectrum is not very sensitive to the presence of nuclei at the highest energies. In the case of our generic composition, the proton component already represents $70\%$ of the total flux at $1.5\,10^{19}$~eV, and a lighter composition could thus provide a similarly good fit of the data. In conclusion, although strong extragalactic magnetic fields would certainly have some impact on the phenomenology of the high-energy CRs, our general results should not be strongly affected.

\subsubsection*{Acknoledgements:} We thank A. G. Mariazzi, T. Dova, A. Watson, D. Heck, S. Sciutto, R. Engel, M. Roth, M. Ave, J. Cronin, T. Yamamoto and G. Pelletier for valuable discussions. This work was supported in part by NSF through grant PHY-04057069 and the KICP under NSF PHY-0114422.


\begin{thebibliography}{}
\bibitem{Kascade} Antoni, T., et al., KASCADE collaboration, 2005, Astropart. Phys., 24, 1, astro-ph/0505413.
\bibitem{Nagano92} Nagano, N., et al.,1992, Nucl. Part. Phys., 18, 423.
\bibitem{Berezinsky+88} Berezinsky,V. S. and Grigorieva,S. I., 1988, Astron. 
Astroph. 199, 1.
\bibitem{Berezinsky+02} Berezinsky, V.~S., Gazizov, A.~Z., \& Grigorieva, S.~I., 2002, astro-ph/0204357.
\bibitem{Berezinsky+04} Berezinsky, V.~S., Grigorieva, S.~I.\& Hnatyk, B.~I, 2004, Astropart. Phys. 21, 617.
\bibitem{HiRescomp} Abbasi, R.U.,  et al., (HiRes), 2005, Ap.J. 622, 910.
\bibitem{Berezinsky+05} Berezinsky, V.~S., Gazizov, A.~Z., \& Grigorieva, S.~I., 2005, Phys. Lett. B, 612, 147.
\bibitem{Allard2005} Allard, D.,  et al., 2005,  A\&A, 443, 29, astro-ph/0505566.
\bibitem{De marco+03} De Marco, D., Blasi, P., Olinto, A.~V.,2003, Astropart. Phys. 20, 53.
\bibitem{Aloisio2005} Aloisio, R. and Berezinsky, V.~S., 2005, astro-ph/0507325.
\bibitem{Lemoine+05} Lemoine, M., 2005, Phys. Rev. D, 71, 083007. 
\bibitem{Gunter} Sigl, G. and Armengaud, E., 2005, astro-ph/0507656.
\bibitem{G66} Greisen, K., 1966, Phys. Rev. Lett. 16, 748.
\bibitem{ZK66}   Zatsepin, G.T. and Kuzmin, V.A.,1966, Sov. Phys. JETP Lett. 4, 78.
\bibitem{PSB} Puget, J.~L., Stecker, F.~W., Bredekamp, J.~H., 1976, ApJ, 205, 638.
\bibitem{Stecker99} Stecker, F.~W., Salamon, M.~H., 1999, ApJ 512, 521.
\bibitem{Anchordoqui1998} Anchordoqui, L., et al., 1998, Phys. Rev D 57, 7103.
\bibitem{Epele+98} Epele, L. and Roulet, E., 1998, astro-ph/9808104.
\bibitem{Bertone+02} Bertone, G., et al., 2002, Phys. Rev. D, 66.
\bibitem{Anchordoqui1999} Anchordoqui, L., et al., 1999, astro-ph/9912081.
\bibitem{Anchordoqui2001} Anchordoqui, L., et al., 2001, Phys. Rev. D 64.
\bibitem{Duvernois96} Du Vernois M~.A and Thayer, M.~R., 1996, ApJ, 465, 982.
\bibitem{MS05}Stecker, F.~W.,   Malkan,  M.~A. and  Scully, S.~T., 2006, ApJ 648, 774, astro-ph/0510449. 
\bibitem{Khan05} Khan, E.~, et al., 2005, Astropart. Phys., 20, 53.
\bibitem{Rachen} Rachen, J.~P., 1996, Interaction processes and statistical properties of the propagation of cosmic-rays in photon backgrounds, PhD thesis of the Bohn University.
\bibitem{Allard2006} Allard, D., et al., 2006, JCAP09(2006)005, astro-ph/0605357.
\bibitem{Bergman+05} Bergman, D. for the HiRes collaboration, 2005, Proceeding of the $29^{th}$ ICRC conference, Pune
(India).
\bibitem{AGASA} Takeda, M., et al., 2003, Proceedings of the $28^{th}$ ICRC (Tsukuba) and Astropart. Phys, 2003, 19,
447.
\bibitem{NaganoWatson} Nagano, M. and Watson, A.~A., 2000, Rev. Mod. Phys. 72, 689.
\bibitem{newAGASA} Teshima, M., 2006, in the Proceedings of CRIS06, Catania, Italy. 
\bibitem{Augerspec} P. Sommers (Pierre Auger Collaboration), 2005, in the Proceedings of  ICRC05, Pune, India.
\bibitem{Parizot+05} Parizot, E., 2005, New views of the Universe, Proceedings of the V$^{th}$ rencontres du Vietnam
(Hanoi 2004), astro-ph/0501274.
\bibitem{Hoerandel03} Hoerandel, J.~R., et al., 2003, Astropart. Phys. 19, 193. 
\bibitem{Dova05} Dova, M.~T., Mariazzi A.~G., Watson A.~A., 2005, Proceeding of the $29^{th}$ ICRC, Pune (India).
\bibitem{Watson} Watson, A.~A., 2004, astro-ph/0410514.
\bibitem{HaverahPark} Ave, M., et al., 2003, Astropart. Phys. 19, 61.
\bibitem{Ave2003b} Ave, M., et al., 2003,  Proceeding of the $28^{th}$ ICRC conference, Tsukuba (Japan).  
\bibitem{PAO} Pierre Auger Observatory Design Report, 1997, Auger Collaboration, Fermi National Accelerator Observatory.
\bibitem{Hoerandel05} Hoerandel, J.~R., et al., 2005, astro-ph/0508015.
\bibitem{Corsika} Heck, D., et al., 1998, Report FZKA 6019, Forschungszentrum Karlshrue.
\bibitem{Aires} Sciutto, S.~J., 2001, Proceedings of the $27^{th}$ ICRC (Hamburg), astro-ph/0106044.
\bibitem{Conex1} Pierog, T., et al., 2004, proceedings of the 13$^{th}$ International Symposium of Very High-Energy Cosmic Ray Interactions at the NESTOR Institute, Pylos (Greece), astro-ph/0411260.
\bibitem{QGSJet} Kalmykov, N.~N. and Ostapchenko, S~., 1993, Phys. At. Nucl. 56, 346.
\bibitem{Sibyll} Engel, R., et al., 1999, Proceedings of the $26^{th}$ ICRC (Salt Lake City).
\bibitem{QGSJetII} Ostapchenko, S., 2004, Proceedings of the $13^{th}$ international symposium of very high energy cosmic-rays interactions (Pylos, Greece), astroph/0412591.
\bibitem{WW} Wibig, T. and Wolfendale, A.~W., 2004, astro-ph/0410624.
\bibitem{Bird+93} Bird, D., et.al, 1993, Phys. Rev. Let. 71, 3401.
\bibitem{Afanasiev93} Afanasiev, B.~N., 1993, Proceedings of the Tokyo workshop on techniques for the study of the
extremely high energy cosmic-rays.
\bibitem{Sokolsky05} Sokolsky, P., for the HiRes collaboration,  2005, Proceedings of the $29^{th}$ ICRC, Pune (India).
\bibitem{HiRes-Mia} Abu-Zayyad, T., et. al.,  (HiRes-Mia), 2000, Phys. Rev. Lett. 84, 4276.
\bibitem{Shinozaki} Shinosaki, K., 2004, Nucl. Phys. Proc. Suppl., 136, 18.
\bibitem{Dawson98} Dawson, B.~R., et al., 1998, Astropart. Phys. 9, 331.
\bibitem{Berezinsky+05b} Berezinsky, V.~S., astro-ph/0509675.

\bibitem{Alvarez+02} Alvarez Muniz, J., et al., 2002, Phys. Rev. D 66, 033011.
\bibitem{YakutskMuons}Knurenko, S.~P., et al., 2005, J. Mod. Phys. A20, 6900.
\bibitem{Grande} Haungs, A., et al. KASCADE-Grande collaboration, 2005, Workshop on Physics of the End of the
Galactic Cosmic Ray Spectrum, astro-ph/0508286.
\bibitem{Sigl05} Sigl, G., Miniati, F. and Ensslin, T., 2004, Phys. Rev. D 70.
\bibitem{Dolag05} Dolag, K., Grasso, D., Springel, V., Tkachev, I, 2004, astro-ph/0410419.
\bibitem{Parizot04} Parizot, E., 2004, Nucl. Phys. B, 136, 169

\end{thebibliography}
\end{document}